\documentclass[aps,prb,showpacs,twocolumn]{revtex4}

\usepackage{graphicx}
\usepackage{bm}
\usepackage{amssymb}
\usepackage{amsmath}

\begin{document}

\title{Conductance calculations for quantum wires and interfaces:\protect\\
mode matching and Green functions}
\author{P. A. Khomyakov} \author{G. Brocks}
\thanks{Corresponding author}
\email[E-mail: ]{g.brocks@tn.utwente.nl} \homepage{http://www.tn.utwente.nl/cms}
\author{V. Karpan} \author{M. Zwierzycki} \author{P. J. Kelly}
\affiliation{Computational Materials Science, Faculty of Science and Technology and MESA+
Research Institute, University of Twente, P. O. Box 217, 7500 AE Enschede, The Netherlands.}
\date{\today}

\begin{abstract}
Landauer's formula relates the conductance of a quantum wire or interface
to transmission probabilities. Total transmission probabilities are
frequently calculated using Green function techniques and an expression
first derived by Caroli. Alternatively, partial transmission
probabilities can be calculated from the scattering wave functions that
are obtained by matching the wave functions in the scattering region to
the Bloch modes of ideal bulk leads. An elegant technique for doing this,
formulated originally by Ando, is here generalized to any Hamiltonian
that can be represented in tight-binding form. A more compact expression
for the transmission matrix elements is derived and it is shown how all
the Green function results can be derived from the mode matching technique.
We illustrate this for a simple model which can be studied analytically,
and for an Fe$|$vacuum$|$Fe tunnel junction which we study using
first-principles calculations.
\end{abstract}


\maketitle

\section{Introduction}
Since the discovery of giant magnetoresistance in metallic multilayers
there has been considerable interest in studying electronic transport
in layered materials.\cite{Baibich:prl88,Binasch:prb89} At the same
time, experimental control of the lateral scale has enabled studies of
electronic transport in quantum wires of atomic
dimensions.\cite{Agrait:prp03} Because of the small dimensions
involved, the transport properties of both of these systems should be
understood on the basis of their atomic structure. This perception
has generated a large effort in recent years to calculate the
conductance of multilayers and quantum wires from first
principles. Several different approaches have been formulated
which have a common basis in the Landauer-B\"{u}ttiker approach or
are equivalent to it. In the linear response regime, the conductance
$\mathcal{G}$ is expressed as a quantum mechanical scattering
problem\cite{Buttiker:prb85} and can be simply related to the total
transmission probability at the Fermi energy, $T(E_F)$, as
\begin{equation}\label{eq1}
\mathcal{G}=\frac{e^2}{h}T(E_F).
\end{equation}
The multilayer or quantum wire is generally considered as a scattering
region of finite size, sandwiched between two semi-infinite ballistic
wires. Aiming at a materials specific description, most current
approaches treat the electronic structure within the framework of density
functional theory (DFT).\cite{Schep:prb97,vanHoof:prb99,Kudrnovsky:prb00,
Xia:prb01,Riedel:prb01,Taylor:prb01,Brandbyge:prb02,Wortmann:prb02b,
Thygesen:prb03,Mavropoulos:prb04}

Frequently the conductance is calculated using a Green function expression
first derived by Caroli {\it et al.}\cite{Caroli:jpc71,Datta:95}
An alternative technique, suitable for Hamiltonians that can be represented
in tight-binding form, has been formulated by Ando.\cite{Ando:prb91} It is
based upon directly matching the wave function in the scattering region
to the Bloch modes of the leads. The latter technique has been applied to
conductance calculations at the empirical tight-binding level,
\cite{Nicolic:prb94} as well as on the first-principles DFT level.
\cite{Xia:prb01,Xia:prb02,Xia:prl02,Zwierzycki:prb03} The relationship
between the mode matching\cite{Ando:prb91} and Green function
\cite{Caroli:jpc71,Datta:95,Krstic:prb02} approaches is not immediately
obvious. Indeed, it was recently stated that Ando's approach is incomplete
and does not yield the correct expression for the conductance.
\cite{Krstic:prb02}

In this paper we demonstrate that the two approaches are completely equivalent. In the Green
function approach, a small imaginary part must be added to or subtracted from the energy in
order to distinguish between the retarded and advanced forms.\cite{Schep:prb97,vanHoof:prb99,
Kudrnovsky:prb00,Taylor:prb01,Brandbyge:prb02,Wortmann:prb02b,
Thygesen:prb03,Mavropoulos:prb04,Krstic:prb02} In mode matching, scattering wave functions
are calculated that incorporate the retarded or advanced boundary conditions directly. This
makes it possible to solve the scattering problem also at real, instead of complex energies.
In addition to yielding the total conductance, by focussing on wave functions the
contribution of each individual scattering channel can be identified. In particular, we
derive a simple, compact expression for the transmission matrix elements, see
Eq.~(\ref{eqtt4}).

The paper is organized as follows. In the next section the Hamiltonian
we will use is introduced. This model allows us to study both quantum
wires that are finite in the directions perpendicular to the wire, and
systems that are periodic in these directions such as single interfaces,
sandwiches and multilayers. We will use the single term ``quantum wire"
to describe both systems. In Secs. \ref{wave} and \ref{green} the mode
matching and Green function techniques are summarized. The equivalence
of the transmission matrices obtained using these two techniques is
demonstrated in Sec. \ref{relation} and the Caroli expression for the
conductance is derived from the mode matching expressions.
In Sec. \ref{examples} the two techniques are applied first to a simple
analytical model,\cite{Sautet:prb88} and then to an Fe$|$vacuum$|$Fe tunnel
junction using numerical first-principles calculations. The main
conclusions are summarized in Sec. \ref{conclusions}.

\section{Hamiltonian}
\begin{figure}[!]
\includegraphics[width=8.5cm,keepaspectratio=true]{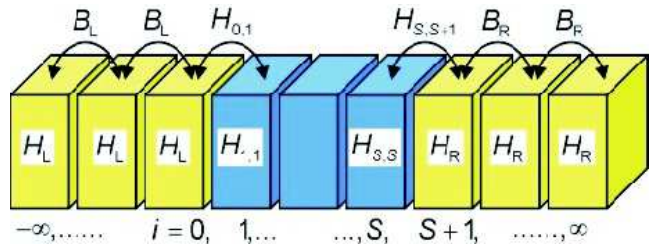}
\caption{(Color online) Hamilton matrix of a quantum wire divided into slices. The left (L)
and right (R) leads are ideal periodic wires that span the cells $i=-\infty ,\ldots, 0$ and
$i=S+1,\ldots,\infty$, respectively. The scattering region spans cells $i=1,\dots,S$.}
\label{fig:hamiltonian}
\end{figure}

We set up a tight-binding representation of the Hamiltonian. This is not
a severe restriction since a first-principles DFT implementation that uses
a representation on an atomic orbital basis set has the same mathematical
structure as a tight-binding model.\cite{Note:ortho_remark,Xia:prb05}
Alternatively, an implementation that uses a representation of the
Hamiltonian on a grid in real space, can also be mapped onto a tight-binding
model.\cite{Khomyakov:prb04} We begin by dividing the system into slices
(``principal layers'') perpendicular to the wire direction.
\cite{MacKinnon:zfp85} The thickness
of these slices is chosen such that there is only an interaction between
neighboring slices. Labelling each slice with an index $i$,
the Schr\"{o}dinger equation of the quantum wire becomes
\begin{equation}
-\mathbf{H}_{i,i-1}\mathbf{c}_{i-1}+\left( E\mathbf{I}-\mathbf{H}%
_{i,i}\right) \mathbf{c}_{i}-\mathbf{H}_{i,i+1}\mathbf{c}_{i+1}=0,
\label{eq2}
\end{equation}
for $i=-\infty ,\ldots ,\infty $. Assuming that each slice contains $N$ sites and/or
orbitals, $\mathbf{c}_{i}$ is a vector of dimension $N$ containing the wave function
coefficients on all sites and/or orbitals of slice $i$. The $N\times N$ matrices
$\mathbf{H}_{i,i}$ and $\mathbf{H}_{i,i\pm1}$ consist of on-slice and hopping matrix elements
of the Hamiltonian, respectively. $\mathbf{I}$ is the $N\times N$ identity matrix. A
schematic representation of the structure of the Hamiltonian is given in
Fig.~\ref{fig:hamiltonian}.

Eq.~(\ref{eq2}) is valid both for quantum wires that are finite in the
directions parallel to the slices, and for layered systems that are
periodic in these directions. In the latter case, translations in the
transverse direction can be described in terms of a Bloch wave vector
in the two-dimensional Brillouin zone, $\mathbf{k}_\parallel$, which
is a good quantum number and the system becomes effectively one
dimensional. Explicit expressions for the Hamilton matrix elements
depend upon the particular localized orbital basis or real space grid
representation used.\cite{Note:singular_remark} Since details
of the tight-binding muffin tin orbital scheme used in
Refs.~\onlinecite{Xia:prb01,Xia:prb02,Xia:prl02,Zwierzycki:prb03}
are given in Ref.~\onlinecite{Xia:prb05} and of the real-space
high-order finite difference method can be found in
Ref.~\onlinecite{Khomyakov:prb04}, they will not be discussed
further here.

The system is divided into three parts, with $i=-\infty ,\ldots ,0$
corresponding to the left lead (L), $i=1,\ldots ,S$ \ to the scattering
region (S) and $i=S+1,\ldots ,\infty $ to the right lead (R). The leads
are assumed to be ideal wires characterized by a periodic potential. It
is then sufficient to identify a slice with a translational period along
the wire. By construction, the Hamilton matrix is the same for each slice
of the leads, i.e.
$\mathbf{H}_{i,i}   \equiv \mathbf{H}_{\mathrm{L/R}}$,
$\mathbf{H}_{i,i-1} \equiv \mathbf{B}_{\mathrm{L/R}}$ and
$\mathbf{H}_{i,i+1} \equiv \mathbf{B}_{\mathrm{L/R}}^{\dagger}$ for the
left/right leads.
Fig.~\ref{fig:hamiltonian} summarizes our model of a quantum wire.

\section{Mode matching approach\label{wave}}
Eq.~(\ref{eq2}) can be solved in a particularly convenient way by a
technique which we call mode matching. In this, we follow Ando and
generalize his approach to a slice geometry.\cite{Ando:prb91}

\subsection{Bloch matrices}\label{blochmatrix}
The first step consists of finding solutions for the leads, for
which Eq.~(\ref{eq2}) can be simplified to
\begin{equation}
-\mathbf{Bc}_{i-1}+
\left( E\mathbf{I}-\mathbf{H}\right) \mathbf{c}_{i}-
\mathbf{B}^{\dagger }\mathbf{c}_{i+1}=0.
\label{eq5}
\end{equation}
These equations hold for $i=-\infty ,\ldots ,-1$ and $i=S+2,\ldots ,\infty $,
i.e. the left and right leads. To keep the notation as simple as possible, we
have omitted the subscripts L and R, see Fig.~\ref{fig:hamiltonian}. Since
the leads are periodic wires, one can make the ansatz that the solutions
have Bloch symmetry, i.e. $\mathbf{c}_{i}=\lambda \mathbf{c}_{i-1}$,
$\mathbf{c}_{i+1}=\lambda ^{2}\mathbf{c}_{i-1}$, where $\lambda $ is the
Bloch factor. Substituting this into Eq.~(\ref{eq5}) results in a quadratic
eigenvalue equation of dimension $N$. The latter can be solved most easily
by transforming it to an equivalent linear (generalized) eigenvalue problem
of dimension $2N$ and solving this by a standard algorithm.
\cite{Tisseur:siam01,Golub:96}

It can be shown that the equation generally has $2N$ solutions, which
can be divided into $N$ right-going modes and $N$ left-going modes,
\cite{Molinari:jpa97} labelled as $(+)$ and $(-)$ in the following.
Right-going modes are either evanescent waves that are decaying to the
right, or waves of constant amplitude that are propagating to the right,
whereas left-going modes are decaying or propagating to the left.
We denote the eigenvalues by $\lambda _{n}(\pm )$ where $n=1,\ldots ,N$,
the corresponding eigenvectors by $\mathbf{u}_{n}(\pm )$ and write the
eigenvalue equation as
\begin{equation}
-\mathbf{Bu}_{n}(\pm )+\left( E\mathbf{I}-\mathbf{H}\right) \lambda _{n}(\pm
)\mathbf{u}_{n}(\pm )-\mathbf{B}^{\dagger }\lambda _{n}(\pm )^{2}\mathbf{u}%
_{n}(\pm )=0.  \label{eq6a}
\end{equation}
In the following we assume that the vectors $\mathbf{u}_{n}(\pm )$ are
normalized; note that in general they are \textit{not} orthogonal.

One can distinguish right- from left-going evanescent modes on the basis
of their eigenvalues; right going evanescent modes have $|\lambda(+)|<1$
and left going evanescent modes have $|\lambda(-)|>1$. Propagating modes
have $|\lambda(\pm)|=1$, so here one has to determine the
Bloch velocity and use its sign to distinguish right from
left propagation. The Bloch velocities are given by the expression
\begin{equation}
v_{n}(\pm)=-\frac{2a}{\hbar }\mathrm{Im}\left[ \lambda _{n}(\pm) \mathbf{u}_{n}(\pm)^{\dagger
}\mathbf{B}^{\dagger } \mathbf{u}_{n}(\pm)\right],  \label{eq19a}
\end{equation}
where $a$ is the slice thickness. A derivation of this equation is given
in Appendix A.

Since the eigenvectors are non-orthogonal, it is convenient to define
dual vectors $\widetilde{\mathbf{u}}_{n}(\pm )$
\begin{equation}
\widetilde{\mathbf{u}}_{n}^{\dagger }(\pm )\mathbf{u}_{m}(\pm )=\delta_{n,m} \: ; \:
\mathbf{u}_{n}^{\dagger }(\pm )\widetilde{\mathbf{u}}_{m}(\pm )=\delta _{n,m}.
\label{eq8}
\end{equation}
Any wave function in the leads can be expressed as a linear combination of the lead modes.
This can be done in a compact form using the two $N\times N$ \textit{Bloch matrices} for
right- and left-going solutions
\begin{equation}
\mathbf{F(\pm )}=\sum_{n=1}^{N}\lambda _{n}(\pm )\mathbf{u}_{n}(\pm )%
\widetilde{\mathbf{u}}_{n}^{\dagger }(\pm ).  \label{eq7}
\end{equation}
For any integer $i$, $\mathbf{F}^{i}$ is given by a similar expression
but with $\lambda _{n}$ replaced by $\lambda _{n}^{i}$. From the
foregoing it is easy to show that the Bloch matrices obey the equation
\begin{equation}
-\mathbf{BF}^{-1}(\pm)+\left( E\mathbf{I}-\mathbf{H}\right) -
\mathbf{B}^{\dagger}\mathbf{F}(\pm)=0.  \label{eq5aa}
\end{equation}

A general solution in the leads can be expressed in terms of a recursion
relation
\begin{eqnarray}
\mathbf{\,c}_{i} &=&\mathbf{\,c}_{i}(+)+\mathbf{\,c}_{i}(-)  \nonumber \\
&=&\mathbf{F}^{i-j}\mathbf{(+)c}_{j}(+)+\mathbf{F}^{i-j}\mathbf{(-)c}_{j}(-).
\label{eq9}
\end{eqnarray}
Fixing the coefficients in one slice then sets the boundary conditions
and determines the solution for the whole lead.

\subsection{Transmission matrix}
The scattering region is defined by $i=1,\ldots ,S$, see
Fig.~\ref{fig:hamiltonian}. Right and left of the scattering regions one
has the recursion relations for the states in the leads from Eq.~(\ref{eq9})
\begin{eqnarray}
\mathbf{c}_{-1} &=&\mathbf{F}_{\mathrm{L}}^{-1}\mathbf{(+)c}_{0}(+)+\mathbf{F%
}_{\mathrm{L}}^{-1}\mathbf{(-)c}_{0}(-)  \nonumber \\
&=&\left[ \mathbf{F}_{\mathrm{L}}^{-1}\mathbf{(+)-\mathbf{F}_{\mathrm{L}%
}^{-1}(-)}\right] \mathbf{c}_{0}(+)+\mathbf{F}_{\mathrm{L}}^{-1}\mathbf{(-)c}%
_{0},  \label{eq10}
\end{eqnarray}
with $\mathbf{\,c}_{0} =\mathbf{\,c}_{0}(+)+\mathbf{\,c}_{0}(-)$ and
\begin{equation}
\mathbf{c}_{S+2}=\mathbf{F}_{\mathrm{R}}\mathbf{(+)c}_{S+1}(+)+
                 \mathbf{F}_{\mathrm{R}}\mathbf{(-)c}_{S+1}(-),
\label{eq11}
\end{equation}
where the subscripts L and R distinguish between the Bloch matrices of
the left and right leads.

The boundary conditions for the scattering problem are set up in the
usual way. The vector $\mathbf{c}_{0}(+)$ is treated as the source,
i.e. as a fixed incoming wave from the left lead. There is no incoming
wave from the right lead, so we set $\mathbf{c}_{S+1}(-)=0$.

Having set the boundary conditions, Eqs.~(\ref{eq10}) and (\ref{eq11})
can be used to rewrite Eq.~(\ref{eq2}) in the region not covered by
Eq.~(\ref{eq5}), i.e. for $i=0,\ldots ,S+1$.
This describes the wave function in the scattering region and the
matching to the solutions in the leads. Eq.~(\ref{eq2}) in this region
is rewritten as
\begin{equation}
-\mathbf{H}'_{i,i-1}\mathbf{c}_{i-1}+\left( E\mathbf{I}-%
\mathbf{H}'_{i,i}\right)
\mathbf{c}_{i}-\mathbf{H}'_{i,i+1}\mathbf{c}_{i+1}=\mathbf{Q}_{i}\mathbf{c}_{0}(+),
\label{eq12}
\end{equation}
with a modified Hamilton matrix defined so
$\mathbf{H}'_{i,j} =\mathbf{H}_{i,j}$
for the elements
$\{i,j=0,1\}$, $\{i=1,\ldots ,S \, ; \, j=i,i \pm 1\}$ and $\{i,j=S+1,S\}$,
but with
\begin{eqnarray}
\mathbf{H}'_{0,0} &=&\mathbf{H}_{\mathrm{L}}+\mathbf{B}_{\mathrm{L}}
\mathbf{F}_{\mathrm{L}}^{-1}\mathbf{(-)} ,  \nonumber \\
\mathbf{H}'_{S+1,S+1} &=&\mathbf{H}_{\mathrm{R}}+\mathbf{B}_{\mathrm{R}}^{\dagger }
\mathbf{F}_{\mathrm{R}}\mathbf{(+)}. \label{eq13}
\end{eqnarray}
$\mathbf{H}'_{0,-1} =0$ and $\mathbf{H}'_{S+1,S+2}=0$, so the modified
scattering region is decoupled from the leads.
On the left-hand side of Eq.~(\ref{eq12}), we have a source term with
\begin{equation}
\mathbf{Q}_{0} =\mathbf{B}_{\mathrm{L}}\left[ \mathbf{F}_{\mathrm{L}}^{-1}%
\mathbf{(+)-\mathbf{F}_{\mathrm{L}}^{-1}(-)}\right], \label{eq14}
\end{equation}
and $\mathbf{Q}_{i} =0$ for $i=1,\ldots ,S+1$. Eq.~(\ref{eq12}) defines
a set of $(S+2) \times N$ linear equations. Because the Hamilton matrix
is block tridiagonal, each block being of dimension $N$, this set of
equations can be solved efficiently using a block Gaussian elimination
scheme.\cite{Golub:96} The total wave function $\mathbf{c}_{i}$ can then
be obtained by back substitution.

The transmission is obtained from the wave function in the right lead
$\mathbf{c}_{S+1}(+)$. In particular, choosing the incoming wave as one
of the modes of the left lead, i.e.
$\mathbf{c}_{0}(+)=\mathbf{u}_{\mathrm{L},m}(+)$, generalized transmission
matrix elements $\tau_{n,m}$ are defined by expanding $\mathbf{c}_{S+1}(+)$
in modes of the right lead
\begin{equation}
\mathbf{c}_{S+1}(+)=\sum_{n=1}^{N}\mathbf{u}_{\mathrm{R},n}(+)\tau_{n,m}.
\label{eq17}
\end{equation}
By letting $\mathbf{c}_{0}(+)$ run over all possible incoming modes of
the left lead $\mathbf{u}_{\mathrm{L},m}(+);\,m=1,\ldots,N$, a full
transmission matrix is obtained.

Matrix elements can be defined for all modes, propagating and evanescent,
but of course only matrix elements where $n,m$ denote propagating modes
contribute to the real physical transmission. These modes can be selected
by making use of their eigenvalues; see the discussion following
Eq.~(\ref{eq6a}). The physical transmission matrix elements are then found
by normalizing with respect to the current\cite{Datta:95}
\begin{equation}
t_{n,m}=\sqrt{\frac{v_{\mathrm{R},n}(+)a_\mathrm{L}}{v_{\mathrm{L},m}(+)a_\mathrm{R}}} \; \;
\tau_{n,m}, \label{eq19}
\end{equation}
where $v_{\mathrm{L},m}(+)$  and $v_{\mathrm{R},n}(+)$ are the Bloch velocities in the
direction of the wire for the right-propagating modes $m$ and $n$ in the left and right
leads, respectively, see Eq.~(\ref{eq19a}); $a_\mathrm{L}$ and $a_\mathrm{R}$ are the slice
thicknesses of left and right leads.\cite{Note:current_remark} The total transmission
probability is given by
\begin{equation}
T(E)=\sum_{n,m}^{(+)}\left| t_{n,m}\right| ^{2}, \label{eqt1}
\end{equation}
and the conductance is given by Eq.~(\ref{eq1}) evaluated at $E=E_F$.

\subsection{Green function matrix}
Solving the set of linear equations Eq.~(\ref{eq12}) directly leads
to the conductance. However, to facilitate a connection to the Green
function approach discussed in the next section, we can formulate the
solution in a slightly different way. A finite Green function matrix
$\mathbf{G}'_{i,j}(z),\,i,j=0,\ldots ,S+1$ can be defined by
\begin{equation}
-\mathbf{H}'_{i,i-1}\mathbf{G}'_{i-1,j}+\left( z \mathbf{I}-\mathbf{H}'_{i,i}\right)
\mathbf{G}'_{i,j} -\mathbf{H}'_{i,i+1}\mathbf{G}'_{i+1,j}= \mathbf{I}\delta _{i,j},
\label{eq15}
\end{equation}
with $z$ complex. Note, however, that the matrices $\mathbf{H}'_{0,0}$ and
$\mathbf{H}'_{S+1,S+1}$ are non-Hermitian and $\mathbf{G}'_{i,j}(E)$ is
also uniquely defined for real energies. The Green function matrix allows
the solution of Eq.~(\ref{eq12}) to be written as
\begin{equation}
\mathbf{c}_{i}=\mathbf{G}'_{i,0}(E) \mathbf{Q}_{0}\mathbf{c}_{0}(+).
\label{eq16}
\end{equation}
As before, the transmission can be extracted at $i=S+1$ and comparison
with Eq.~(\ref{eq17}) gives
\begin{equation}
\tau_{n,m}=\widetilde{\mathbf{u}}_{\mathrm{R},n}^{\dagger}(+)
\mathbf{G}'_{S+1,0}(E)\mathbf{Q}_{0}\mathbf{u}_{\mathrm{L},m}(+). \label{eq18}
\end{equation}
which can be used in Eq.~(\ref{eq19}). The Green function matrix block
$\mathbf{G}'_{S+1,0}(E)$ can be found by solving Eq.~(\ref{eq15}) using a recursive algorithm
that resembles a Gaussian elimination scheme.\cite{Xia:prb05,Godfrin:jp91}

\section{Green function approach\label{green}}
An apparently quite different route to the transmission matrix starts
by defining an infinite Green function matrix $\mathbf{G}_{i,j}(z)$  for
$i,j=-\infty ,\ldots ,\infty$ with respect to the original Hamiltonian of
Eq.~(\ref{eq2}).
\begin{equation}
-\mathbf{H}_{i,i-1}\mathbf{G}_{i-1,j}+\left(
z\mathbf{I}-\mathbf{H} _{i,i}\right) \mathbf{G}_{i,j}
-\mathbf{H}_{i,i+1}\mathbf{G}_{i+1,j}= \mathbf{I}\delta _{i,j}.
\label{eq20}
\end{equation}
Choosing $z=\lim_{\eta \rightarrow 0}(E\pm i\eta)$ defines as usual the
retarded/advanced Green function matrix. We shall use
$\mathbf{G}_{i,j}(E)$ to denote the \textit{retarded} Green function
matrix and $\mathbf{G}_{i,j}^a(E)$ to denote the \textit{advanced}
Green function matrix.

\subsection{Partitioning}
\label{partitioning}
Eq.~(\ref{eq20}) is most conveniently solved by applying
a partitioning technique.\cite{Datta:95,Williams:prb82} It is
straightforward to show that the finite part
$\mathbf{G}_{i,j}(z)$ defined for $i,j=0 ,\ldots ,S+1$ can be
derived from a closed set of equations, similar to Eq.~(\ref{eq20}),
but with $\mathbf{H}_{i,j}$ replaced by $\mathbf{H}_{i,j}''$ where
$\mathbf{H}_{i,j}'' =\mathbf{H}_{i,j}$ for the elements
$\{i,j=0,1\}$,
$\{i=1,\ldots ,S ; \, j=i,i\pm 1 \}$ and
$\{i,j=S+1,S \}$, but with
\begin{eqnarray}
\mathbf{H}_{0,0}''(z) &=&\mathbf{H}_{\mathrm{L}}+\mathbf{B}_{\mathrm{%
L}}\mathbf{g}_{\mathrm{L}}(z)\mathbf{B}_{\mathrm{L}}^{\dagger }, \nonumber \\
\mathbf{H}_{S+1,S+1}''(z) &=&\mathbf{H}_{\mathrm{R}}+\mathbf{B}_{%
\mathrm{R}}^{\dagger }\mathbf{g}_{\mathrm{R}}(z)\mathbf{B}_{\mathrm{R}} . \label{eq21}
\end{eqnarray}
Here $\mathbf{g}_{\mathrm{L}}(z)$ and $\mathbf{g}_{\mathrm{R}}(z)$ are the surface Green
functions of the semi-infinite left and right leads, respectively, which can be calculated
using an iterative technique: denoting $\mathbf{G}_{i,j}^{[n]}(z)$ as the solution of an
equation similar to Eq.~(\ref {eq20}), but with $\mathbf{H}_{i,j}=0$ for $\{i>n \vee j>n\}$,
one can easily derive the right-going recursion relation
\begin{eqnarray}
\left[ z\mathbf{I-H}_{n+1,n+1}-\mathbf{H}_{n+1,n}\mathbf{G}_{n,n}^{[n]}(z)%
\mathbf{H}_{n,n+1}\right] && \nonumber \\  \mathbf{G}_{n+1,n+1}^{[n+1]}(z)=\mathbf{I}. &&
\label{eq22}
\end{eqnarray}
For an ideal wire with $i,j=- \infty,\ldots ,n$,
$\mathbf{G}_{n,n}^{[n]}(z)=\mathbf{g}_{\mathrm{L}}(z)$ should be
independent of $n$ resulting in the following
equation for the surface Green function,
\begin{equation}
\left[ z\mathbf{I-H}_{\mathrm{L}}-\mathbf{B}_{\mathrm{L}}\mathbf{g}_{\mathrm{%
L}}(z)\mathbf{B}_{\mathrm{L}}^{\dagger }\right] \mathbf{g}_{\mathrm{L}}(z)=%
\mathbf{I} . \label{eq23}
\end{equation}
Several iterative algorithms have been formulated for solving this
non-linear matrix equation.\cite{MacKinnon:zfp85,Turek:97,Guinea:prb83}
A similar reasoning based upon a left-going recursion for the right
lead results in an equation for the surface Green function
$\mathbf{g}_{\mathrm{R}}(z)$ of the right lead
\begin{equation}
\left[ z\mathbf{I-H}_{\mathrm{R}}-\mathbf{B}_{\mathrm{R}}^{\dagger }\mathbf{g%
}_{\mathrm{R}}(z)\mathbf{B}_{\mathrm{R}}\right] \mathbf{g}_{\mathrm{R}}(z)=%
\mathbf{I.}  \label{eq24}
\end{equation}
Again, setting $z=E + i\eta $ in Eqs.~(\ref{eq23}) and (\ref{eq24}) defines
the usual retarded surface Green functions $\mathbf{g}_{\mathrm{L/R}}(E)$.
Although we are mainly interested in the physical limit $\lim_{\eta \rightarrow 0}$,
in practice a finite value of $\eta$ is often used in order to make the
iterative algorithms stable.

The quantities
\begin{equation}
\mathbf{\Sigma}_{\mathrm{L}}(E)
=\mathbf{B}_{\mathrm{L}}\mathbf{g}_{\mathrm{L}}(E)\mathbf{B}_{\mathrm{L}}^{\dagger }
 \, ; \,
\mathbf{\Sigma }_{\mathrm{R}}(E)
=\mathbf{B}_{\mathrm{R}}^{\dagger }\mathbf{g}_{\mathrm{R}}(E)\mathbf{B}_{\mathrm{R}},
\label{eq25}
\end{equation}
which appear in Eqs.~(\ref{eq21})-(\ref{eq24}), are called the self-energies of the left and
right leads, respectively.\cite{Datta:95} Once these are obtained, the finite Hamilton matrix
of Eq.~(\ref{eq21}) is constructed and the retarded Green function matrix
$\mathbf{G}_{i,j}(E)$ can be found using a recursive algorithm.\cite{Godfrin:jp91} Using the
lead modes the transmission matrix elements can then be calculated, as will be shown in the
next section, Sec.~\ref{greentransmission}. Alternatively, the total transmission probability
can be expressed in a form that does not require the lead modes explicitly, which is
discussed in Sec. \ref{relation}.

\subsection{Transmission matrix}\label{greentransmission}
The transmission matrix can be obtained from the Green function matrix
of Eq.~(\ref{eq20}). To do this, we adapt a Fisher-Lee type of approach
to our tight-binding formulation.\cite{Fisher:prb81}
Assuming that the unperturbed reference wave function is the Bloch mode
$\mathbf{u}_{\mathrm{L},m}(+)$ that comes in from the left lead, the Lippmann-Schwinger
equation\cite{Messiah:61} in tight-binding form is
\begin{eqnarray}
\mathbf{c}_{i} &=& \mathbf{u}_{\mathrm{L},m,i}(+) + \sum_{j,k} \mathbf{G}_{i,j}
\mathbf{V}_{j,k}\mathbf{u}_{\mathrm{L},m,k}(+) \nonumber \\
&=& \left[\mathbf{F}_{\mathrm{L}}^{i}(+) + \sum_{j,k} \mathbf{G}_{i,j}
\mathbf{V}_{j,k}\mathbf{F}_{\mathrm{L}}^{k}(+) \right] \mathbf{u}_{\mathrm{L},m}(+).
\label{eqg1}
\end{eqnarray}
Here $\mathbf{u}_{\mathrm{L},m,i}(+)$ is the reference wave function
in slice $i$. It obeys Bloch symmetry and
$\mathbf{u}_{\mathrm{L},m}(+) \equiv \mathbf{u}_{\mathrm{L},m,0}(+)$
is the Bloch mode at the origin, see Sec.~\ref{blochmatrix}. The matrix
$\mathbf{V}_{j,k}$ represents the perturbation with respect to the ideal
left lead.

Eq.~(\ref{eqg1}) can be simplified using the Dyson equation, which in
tight-binding form reads
\begin{eqnarray}
\mathbf{G}_{i,0} &=& \mathbf{G}_{i,0}^{(0)} + \sum_{j,k} \mathbf{G}_{i,j}
\mathbf{V}_{j,k}\mathbf{G}_{k,0}^{(0)} \nonumber \\
&=& \left[ \mathbf{F}_{\mathrm{L}}^{i}(+) + \sum_{j,k} \mathbf{G}_{i,j}
\mathbf{V}_{j,k}\mathbf{F}_{\mathrm{L}}^{k}(+) \right] \mathbf{G}_{0,0}^{(0)}, \label{eqg2}
\end{eqnarray}
using Eq.~(\ref{eqr9}). Comparing Eqs.~(\ref{eqg1}) and (\ref{eqg2}) one finds the simple
expression
\begin{equation}
\mathbf{c}_{i} = \mathbf{G}_{i,0}(E)\left[ \mathbf{G}_{0,0}^{(0)}(E)\right] ^{-1}
\mathbf{u}_{\mathrm{L},m}(+). \label{eqg3}
\end{equation}

From the definition of the generalized transmission matrix elements, cf. Eq.~(\ref{eq17}),
one then obtains the expression
\begin{equation}
\tau_{n,m} = \widetilde{\mathbf{u}}_{\mathrm{R},n}^{\dagger }(+) \mathbf{G}_{S+1,0}(E)\left[
\mathbf{G}_{0,0}^{(0)}(E)\right] ^{-1}\mathbf{u}_{\mathrm{L},m}(+). \label{eq27}
\end{equation}
To find $\tau_{n,m}$ one needs to calculate only the Green function matrix blocks
$\mathbf{G}_{S+1,0}(E)$ of the full system and $\mathbf{G}_{0,0}^{(0)}(E)$ of the ideal left
lead. The physical transmission matrix elements and the total transmission probability can
then be obtained from Eqs.~(\ref{eq19}) and (\ref{eqt1}).

\section{mode matching versus Green functions
\label{relation}}
The two seemingly different formalisms introduced in Secs. \ref{wave}
and \ref{green} are in fact closely related. In this section we will
show how all Green function results can be obtained from the mode
matching approach. We begin by expressing the Green function matrices
of ideal wires in terms of the Bloch matrices, $\mathbf{F(\pm )}$.
These expressions are then used to prove that the transmission matrix
elements obtained from the mode matching and Green function approaches,
cf. Eqs.~(\ref{eq18}) and (\ref{eq27}), are in fact identical. After
that, we show that the transmission matrix elements are independent of
the exact positions within the leads that are used to match the leads
to the scattering region, apart from a trivial phase factor. Then we
derive from the mode matching expression for the total transmission
probability the Green function expression known as the Caroli
expression.\cite{Caroli:jpc71} Finally, a more compact expression for
the transmission matrix elements is derived.

\subsection{Green functions of ideal wires in terms of Bloch matrices}
\label{blochterms}
We begin by deriving an expression for the retarded Green function
matrix $\mathbf{G}_{i,j}^{(0)}$ of an \textit{ideal infinite wire} in
terms of its eigenmodes. The columns of such a Green function obey the
equation
\begin{equation}
- \mathbf{B}\mathbf{G}_{i-1,j}^{(0)}
+ \left( E^{+}\mathbf{I}-\mathbf{H}\right) \mathbf{G}_{i,j}^{(0)}
- \mathbf{B}^{\dagger}\mathbf{G}_{i+1,j}^{(0)}
= \mathbf{I}\delta _{i,j},
\label{eqr8}
\end{equation}
where $E^+ = E + i \eta$. For $i\neq j$ the solution is similar to that
of the wave functions, see Eq.~(\ref{eq5}). In addition, the retarded
Green function should consist only of propagating waves that move
outwards from the $\delta$-source and/or evanescent states that decay
away from the source.\cite{Messiah:61} From Eq.~(\ref{eq9}), we have
the \textit{column} recursion relations
\begin{eqnarray}
\mathbf{G}_{i,j}^{(0)}(E) &=&\mathbf{F}^{i-j}\mathbf{(-)G}_{j,j}^{(0)}(E),\,i<j,  \nonumber \\
\mathbf{G}_{i,j}^{(0)}(E) &=&\mathbf{F}^{i-j}\mathbf{(+)G}_{j,j}^{(0)}(E),\,i>j. \label{eqr9}
\end{eqnarray}
The diagonal block $\mathbf{G}_{j,j}^{(0)}(E)$ can now be obtained by
combining Eqs.~(\ref{eqr8}) and (\ref{eqr9}), which gives for $i=j$
\begin{equation}
\left[ -\mathbf{B}\mathbf{F}^{-1}\mathbf{(-)}%
+E^{+}\mathbf{I}-\mathbf{H}-\mathbf{B}^{\dagger }\mathbf{F}\mathbf{(+)}\right]
\mathbf{G}_{j,j}^{(0)}=\mathbf{I.}
\label{eqr10}
\end{equation}
Eliminating $E^{+}\mathbf{I}-\mathbf{H}$ using Eq.~(\ref{eq5aa}) then yields
\begin{equation}
\left[\mathbf{G}_{j,j}^{(0)}(E)\right]^{-1}= \mathbf{B}\left[
\mathbf{F}^{-1}\mathbf{(+)-F}^{-1}\mathbf{(-)}\right], \label{eqr13}
\end{equation}
or the equivalent
\begin{equation}
\left[\mathbf{G}_{j,j}^{(0)}(E)\right]^{-1}= \mathbf{B}^\dagger\left[
\mathbf{F}(-)-\mathbf{F}(+)\right], \label{eqr13aa}
\end{equation}
Eqs.~(\ref{eqr9}) and (\ref{eqr13}) represent the full expression for
the Green function $\mathbf{G}_{i,j}^{(0)}$ of an infinite ideal wire
in terms of the Bloch matrices $\mathbf{F}(\pm )$ and thus in terms of
the eigenmodes. Note that we can set $E^+ = E$ since, in terms of the
modes, the retarded Green function matrix is uniquely defined for real
energies.

The \textit{advanced} Green function matrix $\mathbf{G}_{i,0}^{(0)a} (E)$
can be found from a similar procedure. It should consist of propagating
waves that move towards the source and/or evanescent states that grow
towards the source. One can construct two new Bloch matrices
$\mathbf{F}^a(\pm)$, which are similar to those defined in Eq.~(\ref{eq7}).
In $\mathbf{F}^a(+)$ one collects the modes that are decaying to the
right (growing to the left) and modes that are propagating to the left.
$\mathbf{F}^a(-)$ then contains modes that grow or propagate to the right.
\begin{eqnarray}
\mathbf{F}^a(\pm )&=&\sum_{n=1}^{N}\lambda_{n}^a(\pm )\mathbf{u}_{n}^a(\pm)
\widetilde{\mathbf{u}}_{n}^{a\dagger }(\pm),\;\mathrm{with} \label{eqr13a} \\
\lambda_{n}^a(\pm)&=&\lambda_{n}(\mp ),\,\mathbf{u}_{n}^a(\pm)=\mathbf{u}_{n}(\mp)\;
\mathrm{propagating} \nonumber \\
\lambda_{n}^a(\pm )&=&\lambda_{n}(\pm ),\;\mathbf{u}_{n}^a(\pm )=\mathbf{u}_{n}(\pm )\,
\mathrm{evanescent}. \nonumber
\end{eqnarray}
Using these definitions, expressions for the advanced Green function matrix
are obtained from Eqs.~(\ref{eqr9}) and (\ref{eqr13}) by replacing
$\mathbf{F}(\pm)$ with $\mathbf{F}^a(\pm)$.

From the general relation between retarded and advanced Green functions,
$\mathbf{G}_{i,j}=\left( \mathbf{G}_{j,i}^a\right)^\dagger$, the
following \textit{row} recursion relations can be deduced for the
retarded Green function
\begin{eqnarray}
\mathbf{G}_{i,j}^{(0)}(E) &=&\mathbf{G}_{i,i}^{(0)}(E)
\left[\mathbf{F}^{a\dagger}(+)\right]^{j-i},\,i<j,  \nonumber \\
\mathbf{G}_{i,j}^{(0)}(E) &=&\mathbf{G}_{i,i}^{(0)}(E)
\left[\mathbf{F}^{a\dagger}(-)\right]^{j-i},\,i>j. \label{eqr9a}
\end{eqnarray}

The retarded Green function $\mathbf{G}_{i,0}^{(s)} (E)$ of a
\textit{semi-infinite wire} extending from $i=-\infty ,\ldots ,0$
can be obtained using a similar technique. Instead of Eq.~(\ref{eqr9}),
we get
\begin{equation}
\mathbf{G}_{i,0}^{(s)} (E)=\mathbf{F}^{i}(-)\mathbf{ g}(E),\,i<0.
\label{eqr14}
\end{equation}
where $\mathbf{g}(E)=\mathbf{G}_{0,0}^{(s)}(E)$ is the surface Green
function. Using this in Eq.~(\ref{eqr8}) gives for $i=0$ and $j=0$
and for $i=-1$ and $j=0$, respectively,
\begin{eqnarray}
&& \left[ -\mathbf{B}\mathbf{F}^{-1}\mathbf{(-)}%
+E^{+}\mathbf{I}-\mathbf{H}\right] \mathbf{ g}=\mathbf{I},  \nonumber \\
&& \left[ -\mathbf{B}\mathbf{F}^{-1}\mathbf{(-)}%
+E^{+}\mathbf{I}-\mathbf{H}\right] \mathbf{F}^{-1}(-)\mathbf{ g}=\mathbf{B}^{\dagger
}\mathbf{ g}. \label{eqr15}
\end{eqnarray}
Note that the $\mathbf{B}^{\dagger }$ term is absent in the first
equation since we are dealing with a semi-infinite wire. These two
equations can be easily solved to find an expression for the surface
Green function
\begin{equation}
\mathbf{ g}(E)=\mathbf{F}^{-1}\mathbf{(} - \mathbf{)}\left( \mathbf{B}^{\dagger }\right)
^{-1}. \label{eqr17}
\end{equation}
Eqs.~(\ref{eqr14}) and (\ref{eqr17}) represent the Green function of a
semi-infinite ideal wire extending from $i=-\infty ,\ldots ,0$. In a
similar fashion, one gets for the Green function of a semi-infinite
ideal wire extending from $i=0,\ldots ,\infty $%
\begin{equation}
\mathbf{G}_{i,0}^{(s)} (E)=\mathbf{F}^{i+1}\mathbf{ (+)B}^{-1},\,i\geq 0. \label{eqr18}
\end{equation}
Analogously to Eqs.~(\ref{eqr14})-(\ref{eqr18}), one can define the
advanced Green function matrix $\mathbf{G}_{i,0}^{(s)a} (E)$ in terms
of the Bloch matrices $\mathbf{F}^a(\pm)$. Moreover, since
 $\left[\mathbf{g}^a\right]^\dagger=\mathbf{g}$, we have the following
relation between the Bloch matrices
\begin{equation}
\mathbf{B}^\dagger\mathbf{F}(\pm)=\mathbf{F}^{a\dagger}(\pm)\mathbf{B}. \label{eqr18a}
\end{equation}

\subsection{Equivalence of mode matching and Green function approaches}
The retarded surface Green functions of the left and right leads can be derived from
Eqs.~(\ref{eqr17}) and (\ref{eqr18})
\begin{equation}
\mathbf{g}_{\mathrm{L}}(E)
=\mathbf{F}_{\mathrm{L}}^{-1}(-)\left[\mathbf{B}_{\mathrm{L}}^{\dagger }\right]^{-1}
\, ; \,
\mathbf{g}_{\mathrm{R}}(E)
=\mathbf{F}_{\mathrm{R}}(+)\mathbf{B}_{\mathrm{R}}^{-1}.
\label{eqr20}
\end{equation}
The retarded self-energies of Eq.~(\ref{eq25}) are then given by
\begin{equation}
\mathbf{\Sigma }_{\mathrm{L}}(E)
=\mathbf{B}_{\mathrm{L}} \mathbf{F}_{\mathrm{L}}^{-1}(-)
\, ; \,
\mathbf{\Sigma }_{\mathrm{R}}(E)
=\mathbf{B}_{\mathrm{R}}^{\dagger }\mathbf{F}_{\mathrm{R}}(+).
\label{eqr6}
\end{equation}
Comparing Eqs.~(\ref{eq13}) and (\ref{eq21}) then establishes
\begin{equation}
\mathbf{G}'_{i,j}(E)=\mathbf{G}_{i,j}(E). \label{eqr7}
\end{equation}
In other words, the two Green functions discussed in Secs. \ref{wave}
and \ref{green} are identical.

By comparing Eqs.~(\ref{eq14}) and (\ref{eqr13}) one has
\begin{equation}
\mathbf{Q}_{0}=\left[ \mathbf{G}_{0,0}^{(0)}(E)\right] ^{-1}. \label{eqr19}
\end{equation}
In conclusion, the two expressions for the generalized transmission
matrix elements, Eqs.~(\ref{eq18}) and (\ref{eq27}), are identical.

\subsection{Invariance of transmission probability
\label{invariance}}
Apart from trivial phase factors, the transmission matrix elements
should not depend on where in the ideal lead the wave function matching
is carried out. In a recent paper it was stated that Ando's expression
for $t_{n,m}$, Eqs.~(\ref {eq18}) and (\ref {eq19}), lacks this invariance
property and is therefore incomplete.\cite{Krstic:prb02} One can however
prove directly from Eq.~(\ref {eq18}) that the transmission matrix
elements do have the required invariance property.\cite{Khomyakov:unpub}
The proof becomes easier if the equivalence of Eqs.~(\ref{eq18}) and
(\ref{eq27}), established above, is used.

The scattering region runs from slices $0$ to $S+1$ if we include the
boundaries with the left and right leads. This means that the Green
function matrix $\mathbf{G}_{i,j}$ with indices $i,j$ outside this
region obeys the equations for the ideal leads. From the column and
row recursion relations, Eqs.~(\ref{eqr9}) and (\ref{eqr9a}), one
derives
\begin{equation}
\mathbf{G}_{S+1+i,j}(E) = \mathbf{F}_{\mathrm{R}}^{i}\mathbf{(+)G}%
_{S+1,0}(E)\left[ \mathbf{F}_{\mathrm{L}}^{a\dagger }\mathbf{(-)}\right] ^{j}, \label{eqr21a}
\end{equation}
for $j<0,\,i>0$. In a similar way, one derives for the Green function
matrix of the left lead
\begin{eqnarray}
\mathbf{G}_{j,j}^{(0)}(E) &=&\mathbf{F}_{\mathrm{L}}^{j}\mathbf{(+)G}_{0,0}^{(0)}(E)\left[
\mathbf{F}_{\mathrm{L}}^{a\dagger }\mathbf{(-)}\right] ^{j}, \label{eqr21}
\end{eqnarray}
for $j<0$.

We now artificially extend the scattering region by including slices
from the left and right leads and let it run from $j<0$ to $S+1+i$
with $i>0$.
Eq.~(\ref{eq27}) gives for the transmission matrix element
\begin{equation}
\tau_{n,m}^{\prime}=\widetilde{\mathbf{u}}_{\mathrm{R},n}^{\dagger
}(+)\mathbf{G}_{S+1+i,j}(E)\left[ \mathbf{G}_{j,j}^{(0)}(E)\right] ^{-1}%
\mathbf{u}_{\mathrm{L},m}(+)  \label{eqr22}
\end{equation}
Using (\ref{eqr21a}) and (\ref{eqr21}) then gives
\begin{eqnarray}
\tau_{n,m}^{\prime} &=&\widetilde{\mathbf{u}}_{\mathrm{R}%
,n}^{\dagger }(+)\mathbf{F}_{\mathrm{R}}^{i}\mathbf{(+)G}_{S+1,0}(E)
\nonumber \\%
&&\left[ \mathbf{G}_{0,0}^{(0)}(E)\right] ^{-1}\mathbf{F}_{\mathrm{L}}^{-j}%
\mathbf{(+)u}_{\mathrm{L},m}(+)  \nonumber \\
&=&\lambda _{\mathrm{R},n}^{i}(+)\lambda _{\mathrm{L},m}^{-j}(+)\tau_{n,m} \nonumber \\
&=&e^{i\alpha }\tau_{n,m}, \label{eqr23}
\end{eqnarray}
with $\alpha$ real. The third line follows from applying Eq.~(\ref{eq7}). The last line
follows from the fact that we are only interested in propagating states and for propagating
states $\left| \lambda \right| =1$. Using this result in (\ref {eq19}) and (\ref {eqt1})
proves the invariance of the total transmission probability with respect to moving the
boundaries between leads and scattering region into the leads.

\subsection{The Caroli expression
\label{transmission}}
The total transmission probability is given by Eq.~(\ref{eqt1}), where
the sum has to be over propagating states only. We can extend the
summation to include all $N$ states (propagating and evanescent) by
defining an $N\times N$ transmission matrix
\begin{equation}
\mathbf{t}=\mathbf{V}_{\mathrm{R}}^{\frac{1}{2}}(+)
\; \mathbf{\tau} \;
\widetilde{\mathbf{V}}_{\mathrm{L}}^{\frac{1}{2}}(+),  \label{eqt2}
\end{equation}
where $\mathbf{\tau}$ is the matrix whose elements are given by Eq.~(\ref{eq18}).
$\mathbf{V}_{\mathrm{R}}(+)$ is defined as the singular, diagonal matrix that has the
velocities $v_{\mathrm{R},n}$ times the constant $\hbar/a_\mathrm{R}$ on the diagonal for the
right-propagating states and zeros for evanescent states. We call it the velocity matrix.
Likewise a pseudo-inverse velocity matrix $\widetilde{\mathbf{V}}_{\mathrm{L}}(+)$ can be
defined, which has $1/v_{\mathrm{L},n} \times a_\mathrm{L}/\hbar$ on the diagonal for
left-propagating states and all other matrix elements are zero. These velocity matrices
project onto the space of the propagating states so that the transmission matrix has only
non-zero values between propagating states. (\ref{eqt1}) can then be expressed in the
familiar form
\begin{eqnarray}
T &=&\mathrm{Tr}\left[ \mathbf{t}^{\dagger }\mathbf{t}\right]   \label{eqt3} \\
&=&\mathrm{Tr}\left[ \mathbf{\tau}^{\dagger
}\mathbf{V}_{\mathrm{R}}(+)\mathbf{\tau}\widetilde{\mathbf{V}}_{\mathrm{L}}(+)\right] .
\nonumber
\end{eqnarray}
Using the above definition of the velocity matrix and Ando's expressions
for the transmission matrix elements, we will show how (\ref{eqt3})
can be rewritten as
\begin{equation}
T=\mathrm{Tr}\left[ \mathbf{\Gamma }_{\mathrm{R}}\mathbf{G}^{r}\mathbf{\Gamma
}_{\mathrm{L}}\mathbf{G}^{a}\right] ,  \label{eqt4}
\end{equation}
where $\mathbf{G}^{r},\mathbf{G}^{a}$ are short-hand notations for $\mathbf{G}_{S+1,0}(E)$
and $\mathbf{G}_{0,S+1}^a(E)$, respectively. The matrices $\mathbf{\Gamma }_{\mathrm{L/R}}$
are defined as
\begin{equation}
\mathbf{\Gamma }_{\mathrm{L/R}}=i\left[ \mathbf{\Sigma }_{\mathrm{L/R}}-\mathbf{\Sigma
}_{\mathrm{L/R}}^\dagger\right] .  \label{eqt5}
\end{equation}
Eq.~(\ref{eqt4}) is known as the Caroli expression,\cite{Caroli:jpc71}
and it is often used to calculate transmission probabilities.
\cite{Datta:95,Krstic:prb02}
It is equivalent to the Kubo-Greenwood expression for the linear response
regime.\cite{Mavropoulos:prb04,Kostyrko:prb00}

The first step is to construct $N\times N$ matrices
$\mathbf{U}(\pm )$, the columns of which are the eigenmodes
$\mathbf{u}_{n}(\pm )$, and diagonal matrices $\mathbf{\Lambda }(\pm )$,
the elements of which are the eigenvalues $\lambda _{n}(\pm )$%
\begin{equation}
\mathbf{U}(\pm )=\left(
\begin{array}{cccc}
\mathbf{u}_{1}(\pm ) & \mathbf{u}_{2}(\pm ) & \ldots  & \mathbf{u}_{N}(\pm )
\end{array}
\right),\label{eqt6}
\end{equation}
\begin{equation}
\mathbf{\Lambda }(\pm )_{m,n}=\lambda _{n}(\pm )\delta _{m,n}. \label{eqt7}
\end{equation}
From Eqs.~(\ref{eq8}) and (\ref{eq9}) it is then easy to show that the dual vectors
$\widetilde{\mathbf{u}}_{n}(\pm )$ form the columns of the matrix $\left[ \mathbf{U}(\pm
)^{-1}\right] ^{\dagger }$ and that the $\mathbf{F}(\pm )$ matrices obey the equation
\begin{equation}
\mathbf{F}(\pm )\mathbf{U}(\pm )=\mathbf{U}(\pm )\mathbf{\Lambda }(\pm ). \label{eqt8}
\end{equation}
In a similar way matrices $\mathbf{U}^a(\pm )$ and $\mathbf{\Lambda }^a(\pm )$ can be
constructed, see Eq.~(\ref{eqr13a}).

Using these definitions, one can generalize the $\mathbf{\tau}$-matrix
of Eq. (\ref{eq18}) to
\begin{equation}
\mathbf{\tau}=\mathbf{U}_{\mathrm{R}}^{-1}(+)\mathbf{G}^r\mathbf{Q}_0
\mathbf{U}_{\mathrm{L}}(+). \label{eqt8a}
\end{equation}
Note that $\mathbf{\tau}$-matrix elements are defined not only between
propagating states, but also between evanescent states. However, as we
remarked above already, only propagating states contribute to the
physical transmission.

The second step is to express the velocity matrices in terms of the
$\mathbf{\Gamma}$-matrices. To do this we use an expression for the
velocity matrix,
\begin{equation}
\mathbf{V(\pm )}=i\left[ \mathbf{U}^{\dagger }(\pm )\mathbf{B}^{\dagger }%
\mathbf{U}(\pm )\mathbf{\Lambda }(\pm )-\mathbf{\Lambda }^{\dagger }(\pm )%
\mathbf{U}^{\dagger }(\pm )\mathbf{BU}(\pm )\right] ,  \label{eqv1}
\end{equation}
which can be shown (see Appendix A for a proof) to be equivalent to the
definition introduced in the first paragraph of this section. Using
\eqref{eqt8}, this can be rewritten for the right lead as
\begin{eqnarray}
\mathbf{V_{\mathrm{R}}(+)} &=&i\mathbf{U}_{\mathrm{R}}^{\dagger }(+) \left[
\mathbf{B}_{\mathrm{R}}^{\dagger }\mathbf{F}_{\mathrm{R}}(+
)-\mathbf{F}_{\mathrm{R}}^{\dagger }(+)\mathbf{B}_{\mathrm{R}}\right]
\mathbf{U}_{\mathrm{R}}(+ )  \nonumber \\
&=&i\mathbf{U}_{\mathrm{R}}^{\dagger }(+)\left[ \mathbf{\Sigma }_{\mathrm{ R}}-\mathbf{\Sigma
}_{\mathrm{R}}^\dagger\right] \mathbf{U}_{\mathrm{R}}(+
)  \nonumber \\
&=&\mathbf{U}_{\mathrm{R}}^{\dagger }(+)\mathbf{\Gamma }_{\mathrm{R}}
\mathbf{U}_{\mathrm{R}}(+).  \label{eqt9}
\end{eqnarray}
The second line follows from (\ref{eqr6}).
A similar relation between the $\mathbf{\Gamma }$-matrix and the
velocity matrix for the left lead can be shown to exist,
\begin{equation}
\mathbf{V_{\mathrm{L}}(+)} =\mathbf{U}_{\mathrm{L}}^{a\dagger }(-)\mathbf{\Gamma
}_{\mathrm{L}} \mathbf{U}^a_{\mathrm{L}}(-),  \label{eqt9a}
\end{equation}
by using (\ref{eqr18a}) and an equivalent expression for the
velocity matrix, Eq.~(\ref{eqv6}). Eqs.~(\ref{eqt9}) and (\ref{eqt9a})
imply that the $\mathbf{\Gamma }$-matrices project onto the space
spanned by the propagating states.

The third step is to introduce a matrix $\mathbf{P}$ that explicitly
projects onto the propagating states of the left lead
\begin{eqnarray}
\mathbf{P}&=&\mathbf{U}_\mathrm{L}(+)\mathbf{I}_p
\left[\mathbf{U}^a_\mathrm{L}(-)\right]^{-1} \nonumber \\
&=& \sum_{n=1}^{N_p}\mathbf{u}_{\mathrm{L},n}(+)
\widetilde{\mathbf{u}}_{\mathrm{L},n}^{a\dagger}(-), \label{eqt10a}
\end{eqnarray}
where the $\mathbf{I}_p$-matrix contains 1 on the $N_p$ diagonal elements that correspond to
propagating states, and 0 at all other positions. Given this projector matrix, it is possible
to prove that
\begin{equation}
\mathbf{Q}_0 \mathbf{P} = i\mathbf{\Gamma}_\mathrm{L}. \label{eqt10b}
\end{equation}
The proof is given in Appendix B. Using this property one has
\begin{eqnarray}
&&\mathbf{Q}_0\mathbf{U}_{\mathrm{L}}(+)\widetilde{\mathbf{V}}_{\mathrm{L}}(+) =
\mathbf{Q}_0\mathbf{P}\mathbf{U}^a_{\mathrm{L}}(-)\widetilde{\mathbf{V}}_{\mathrm{L}}(+)
\nonumber \\
&&=i\mathbf{\Gamma}_\mathrm{L}\mathbf{U}^a_{\mathrm{L}}(-)\widetilde{\mathbf{V}}_{\mathrm{L}}(+)
= i\left[\mathbf{U}^{a\dagger}_{\mathrm{L}}(-)\right]^{-1}
\mathbf{V}_{\mathrm{L}}(+)\widetilde{\mathbf{V}}_{\mathrm{L}}(+) \nonumber \\
&&=i\left[\mathbf{U}^{a\dagger}_{\mathrm{L}}(-)\right]^{-1}\mathbf{I}_p. \label{eqt10c}
\end{eqnarray}
Substituting (\ref{eqt8a}), (\ref{eqt9}) and (\ref{eqt10c}) into
(\ref{eqt3}) leads directly to the Caroli expression, Eq.~(\ref{eqt4}).

\subsection{Transmission matrix: a compact expression}
Using the results of the previous section it is possible to derive a more compact expression
for the transmission matrix. Combining (\ref{eqt2}), (\ref{eqt8a}) and (\ref{eqt10a})
one has
\begin{equation}
\mathbf{t}= \mathbf{V}_{\mathrm{R}}^{\frac{1}{2}}(+)\mathbf{U}_{\mathrm{R}}^{-1}(+)
\mathbf{G}^r\mathbf{Q}_0 \mathbf{P} \mathbf{U}^a_{\mathrm{L}}(-)
\widetilde{\mathbf{V}}_{\mathrm{L}}^{\frac{1}{2}}(+). \label{eqtt1}
\end{equation}
By following the same steps as in Eq.~(\ref{eqt10c}) and using $\mathbf{V}_\mathrm{L}
\widetilde{\mathbf{V}}_{\mathrm{L}}^{\frac{1}{2}}=\mathbf{V}_\mathrm{L}^{\frac{1}{2}}$ this
can be simplified to
\begin{equation}
\mathbf{t}=i\mathbf{V}_{\mathrm{R}}^{\frac{1}{2}}(+)\mathbf{U}_{\mathrm{R}}^{-1}(+)
\mathbf{G}^r
[\mathbf{U}_\mathrm{L}^{a\dagger}(-)]^{-1}\mathbf{V}_\mathrm{L}^{\frac{1}{2}}(+).
\label{eqtt3}
\end{equation}

Writing out the transmission matrix elements gives the compact expression
\begin{equation}
t_{n,m}=i\hbar\sqrt{\frac{v_{\mathrm{R},n}v_{\mathrm{L},m}}{a_\mathrm{R}a_\mathrm{L}}}\;
\widetilde{\mathbf{u}}_{\mathrm{R},n}^\dagger(+) \mathbf{G}_{S+1,0}(E)
\widetilde{\mathbf{u}}^a_{\mathrm{L},m}(-). \label{eqtt4}
\end{equation}
This is the tight-binding equivalent of the Fisher-Lee expression relating transmission and
Green function matrices.\cite{Fisher:prb81}

\section{Examples\label{examples}}
\subsection{A simple analytical model}
\begin{figure}[!]
\includegraphics[width=7.0cm,keepaspectratio=true]{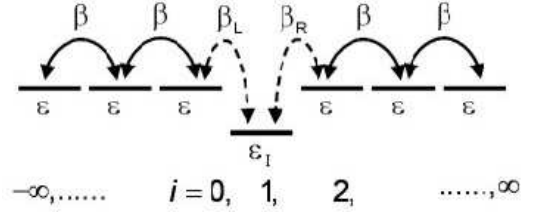}
\caption{The parameters of a one-band nearest neighbor tight-binding model for a single
impurity in a onedimensional chain.} \label{fig:onedmodel}
\end{figure}

We consider a system consisting of a single impurity in a one-dimensional chain and treat
this within a one-band nearest neighbor tight-binding model. The parameters of this model are
given in Fig.~\ref{fig:onedmodel}. This model can be solved analytically,\cite{Sautet:prb88}
so it can serve as a simple test to illustrate the equivalence of the different approaches.

We will first solve the problem using the mode matching approach of Sec.~\ref{wave}. It is
convenient to define a scaled energy by
\begin{equation}
\omega \equiv \frac{E-\epsilon}{2\beta}. \label{eqm1a}
\end{equation}
The model involves only one channel and with $\mathbf{u}_{m}(\pm )=1$ Eq.~(\ref{eq6a})
reduces to
\begin{equation}
-\beta + (E-\epsilon)\lambda(\pm) - \beta \lambda(\pm)^2 = 0, \label{eqm1}
\end{equation}
The roots $\lambda(\pm)$ can be given a more familiar form. For $|\omega| \leq 1$ we define a
wave number $k$ by
\begin{equation}
\cos(ka) = \omega, \label{eqm2a}
\end{equation}
where $a$ is the lattice parameter. From Eqs.~(\ref{eqm1}), (\ref{eqm2a}) one then obtains
\begin{equation}
\lambda(\pm)=e^{\pm ika}, \label{eqm2aa}
\end{equation}
which describes propagating states. For $|\omega| > 1$ one defines $\kappa$ by
\begin{equation}
\cosh(\kappa a) = \left|\omega\right|. \label{eqm2b}
\end{equation}
One obtains $\lambda(\pm)=\exp(\mp \kappa a)$ if $\omega > 1$ and $\lambda(\pm)=-\exp(\mp
\kappa a)$ if $\omega < -1$; both cases describe evanescent states.

Since the scattering region consists of a single impurity, $S=1$,
Eqs.~(\ref{eq12})-(\ref{eq14}) give three linear equations with three unknowns describing the
scattering problem. In a one-channel model one has $\mathbf{F}(\pm )=\lambda(\pm)$ and
$\lambda(\pm)^{-1}=\lambda(\mp)$. There is only one possible incoming wave, so
$\mathbf{c}_0(+)=1$. The linear equations then become in matrix form
\begin{equation}
\begin{array}{c}
\,        \\
\mathbf{A} \\
\,
\end{array}
\left(\begin{array}{c} c_0 \\ c_1 \\ c_2\end{array}\right) =
\left(\begin{array}{c}\beta[\lambda(-)-\lambda(+)] \\
0 \\ 0 \end{array}\right), \label{eqm3}
\end{equation}
with
\begin{equation}
\begin{array}{c}
\,         \\
\mathbf{A} = \\
\,
\end{array}
\left(\begin{array}{ccc}
E-\epsilon-\beta\lambda(+) & -\beta_{\mathrm{L}}     & 0 \\
-\beta_{\mathrm{L}}     & E-\epsilon_{\mathrm{I}} &
-\beta_{\mathrm{R}} \\
0                               & -\beta_{\mathrm{R}}     & E-\epsilon-\beta\lambda(+)
\end{array}\right). \label{eqm3a}
\end{equation}
Solving this set of equations and using Eqs.~(\ref{eqm2a}) and (\ref{eqm2aa}) we obtain the
compact expression
\begin{equation}
c_2 = e^{2ika} \frac{-i f \sin(ka)} {d+ (1-b) \cos(ka) - i b \sin(ka)}, \label{eqm4}
\end{equation}
defining the dimensionless parameters
\begin{equation}
b=\frac{\beta_{\mathrm{L}}^2+\beta_{\mathrm{R}}^2}{2\beta^2},\;
d=\frac{\epsilon-\epsilon_{\mathrm{I}}}{2\beta},\;
f=\frac{\beta_{\mathrm{L}}\beta_{\mathrm{R}}}{\beta^2}. \label{eqm5}
\end{equation}
Applying Eqs.~(\ref{eq17})-(\ref{eqt1}) then yields for the total transmission probability
\begin{equation}
T(E)=\left|c_2\right|^2. \label{eqm6}
\end{equation}
Using Eqs.~(\ref{eqm2a}) and (\ref{eqm4}) it is easy to show that this transmission
probability is identical to Eq.~(15) of Ref.~\onlinecite{Sautet:prb88}, which was obtained
using a different technique.

It is instructive to solve the same problem using the Green function approach of
Sec.~\ref{partitioning}. First one has to find the surface Green functions of the leads from
Eqs.~(\ref{eq23}) and (\ref{eq24}), which for the current model become
\begin{equation}
\left[E-\epsilon-\beta^2g(E)\right]g(E)=1, \label{eqn1}
\end{equation}
where $E$ is a real energy. This equation has the solutions
\begin{equation}
g^\pm(E)=\frac{e^{\pm ika}}{\beta}, \label{eqn2}
\end{equation}
for both leads. The Green function matrix in the scattering region can then be found from
Eqs.~(\ref{eq20}) and (\ref{eq21}), which can be combined in the $3 \times 3$ matrix equation
\begin{equation}
\mathbf{A}\mathbf{G}(E)=\mathbf{I}, \label{eqn3}
\end{equation}
where $G_{i,j}(E)$, $i,j=0,\ldots,2$ are the matrix elements of $\mathbf{G}(E)$ and
$\mathbf{A}$ is given by Eq.~(\ref{eqm3a}). Inverting $\mathbf{A}$ yields the matrix element
\begin{equation}
G_{2,0}(E)=\frac{f}{2\beta} \frac{e^{2ika}}{d+ (1-b) \cos(ka) - i b \sin(ka)}, \label{eqn5}
\end{equation}
with the parameters $b$, $d$ and $f$ defined by Eq.~(\ref{eqm5}). The relevant Green function
matrix element for the ideal lead is found from Eq.~(\ref{eqr13})
\begin{equation}
G_{0,0}^{(0)}(E)=\frac{i}{2\beta\sin(ka)}. \label{eqn4}
\end{equation}
Using these results in Eq.~(\ref{eq27}) one observes that the expression for the (one
channel) transmission matrix element becomes identical to Eq.~(\ref{eqm4}).

Finally one can calculate the transmission probability from the Caroli expression given in
Sec.~\ref{transmission}, cf. Eq.~(\ref{eqt4}). Using Eqs.~(\ref{eq25}), (\ref{eqt5}) and
(\ref{eqn2}) one obtains
\begin{equation}
\Gamma_\mathrm{L}= \Gamma_\mathrm{R}= -2\beta \sin(ka), \label{eqn6}
\end{equation}
for left and right leads. Using Eqs.~(\ref{eqn5}), (\ref{eqn6}) and
$G_{0,2}^a=\left(G_{2,0}\right)^*$ in Eq.~(\ref{eqt4}) then yields an expression for the
transmission probability that is identical to Eq.~(\ref{eqm6}). It illustrates the
equivalence of the different approaches for calculating the transmission in this simple
model.

In addition to providing a channel for propagating states, an impurity can also give rise to
localized states, whose energy is outside the energy band of the chain, cf.
Eq.~(\ref{eqm2b}). Such a state does not contribute to the physical transmission, but the
transmission amplitude has a pole at an energy that corresponds to a localized
state.\cite{Landau:81} Within the mode matching approach this corresponds to an energy at
which $c_{S+1}$ becomes infinite. For the present model the energies of localized states can
be obtained by setting $k=i\kappa$ and setting the denominator to zero in Eq.~(\ref{eqm4}).
This leads to the equation
\begin{equation}
\left(\omega+d\right)\left(\omega+\mathrm{sgn}(\omega)\sqrt{\omega^2-1}\right)-b=0; \;
|\omega|>1, \label{eqm7}
\end{equation}
the roots of which give the energies of the localized states. Again these results are
equivalent to the results obtained using the approach of Ref.~\onlinecite{Sautet:prb88}.

Within the Green function approach the energies of the localized states are given by the
poles of the Green function matrix. Via Eq.~(\ref{eqn5}) this again leads to
Eq.~(\ref{eqm7}). Alternatively, since the Green function matrix is the inverse of the
$\mathbf{A}$-matrix, cf. Eq.~(\ref{eqn3}), its poles are given by the roots of
$\det(\mathbf{A})=0$. This equation is equivalent to Eq.~(\ref{eqm7}), as is easily shown by
setting $\lambda(+)=\pm \exp(-\kappa a)$ in the $\mathbf{A}$-matrix.

\subsection{Fe$|$vacuum$|$Fe tunnel junction}
As an example of a more complex system, we consider an Fe$|$vacuum$|$Fe
tunnel junction where the electronic structure is treated using the
local density approximation of DFT.\cite{Perdew:prb81} The calculations
are based upon a tight-binding muffin tin orbital (TB-MTO) atomic spheres
approximation (ASA) implementation
\cite{Xia:prb01,Xia:prb02,Xia:prl02,Zwierzycki:prb03,Xia:prb05} of the
formalism described in Sect.~\ref{wave}.

The first step in the calculation is the self-consistent determination
of the electronic structure of the tunnel junction using the layer
Green function approach of Ref.~\onlinecite{Turek:97}. The Fe leads are
oriented in the (001) direction and the atoms at the Fe(001) surfaces
are kept at their unrelaxed bulk positions. For the bcc structure and
TB-MTOs,\cite{Andersen:85} a principal layer in the (001) direction
contains two monolayers of Fe with a thickness of 2.866 \AA. The vacuum
region is modeled by a number of such slices, of the same thickness,
filled with ``empty'' spheres of the same size and packing as the Fe
atomic spheres. The atomic sphere potentials of the vacuum region and
four monolayers (two principal layers) of Fe on either side of the
vacuum are calculated self-consistently while the potentials of more
distant layers are kept at their bulk values. These potentials then
form the input to a transmission calculation based on mode matching.
\cite{Xia:prb01,Xia:prl02,Zwierzycki:prb03} Further technical details
can be found in Ref.~\onlinecite{Xia:prb05}.

A useful quantity to extract from the self-consistent layer calculation
is the layer density of states (LDOS) $\rho_i(E)$. It is related to the
retarded Green function matrix defined in Eq.~(\ref{eq20}) by
\begin{equation}
\rho_i(E+i\eta)=-\pi^{-1}\mathrm{Im}\mathrm{Tr}[\mathbf{G}_{i,i}(E+i\eta)]
\label{eqfe1}
\end{equation}
where the trace refers to the usual $lm$ angular momentum indices
characterizing MTOs. For reasons of numerical stability the retarded
Green function matrix is calculated by adding a finite imaginary part
to the energy; we have used $\eta=0.0025$ Ry. In the mode matching
approach, the LDOS can be directly expressed in terms of the wave
functions or, alternatively, in terms of the Green function matrix of
Eq.~(\ref{eq15})
\begin{equation}
\rho_i(E)=-\pi^{-1}\mathrm{Im}\mathrm{Tr}[\mathbf{G}'_{i,i}(E)] .
\label{eqfe2}
\end{equation}
This Green function matrix can be calculated for a real energy, but in
order to make a comparison to the results obtained with Eq.~(\ref{eqfe1}),
we add an imaginary part, $\eta=0.0025$ Ry.

\begin{figure}[!]
\includegraphics[width=8.5cm,keepaspectratio=true]{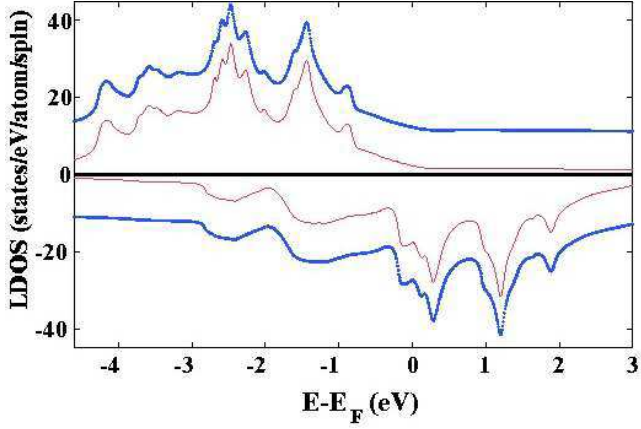}
\caption{(Color online) Layer density of states (LDOS) at the
Fe(001) surface layer. Top panel: majority spin. Top curve (red,
dashed): as calculated using the layer Green function
method.\cite{Turek:97} Bottom curve (blue, solid): as calculated
using the mode matching approach. Both calculations use
$\eta=0.0025$ Ry. For clarity, the top curve has been displaced by
10 units along the $y$-axis. Bottom panel: minority spin, the LDOS
is shown with a negative sign.} \label{fig:LDOS}
\end{figure}

In Fig.~\ref{fig:LDOS} we compare the LDOS obtained using \eqref{eqfe1}
and \eqref{eqfe2} for the topmost Fe monolayer of (001) Fe$|$vacuum$|$Fe
where the vacuum layer was so thick (four principal layers, corresponding
to a thickness of 11.466 \AA) that the LDOS corresponds closely to that
of a free Fe(001) surface. The two curves, displaced vertically for
clarity, are indistinguishable, illustrating the equivalence of the two
Green functions defined in \eqref{eq15} and \eqref{eq20}. Moreover, the
LDOS are in essential agreement with results found previously for a
Fe(001) surface.\cite{Wang:prb81}

If one resolves the LDOS into contributions from different parts
of the surface Brillouin zone, then the contribution at
$\overline{\Gamma}$ ($\mathbf{k}_\parallel=0$) exhibits sharp
peaks near the Fermi level. These are associated with
characteristic surface states found on (001) surfaces of bcc
transition metals.\cite{Stroscio:prl95} These surface states are
mainly derived from $d_{3z^2-r^2}$ orbitals on the surface atoms
projecting into the vacuum, and they have $\Delta_1$ symmetry.
\cite{Uiberacker:prb01,Uiberacker:prb02} They become most clearly
visible if one filters out the contribution to the LDOS at
$\overline{\Gamma}$ of the states with $\Delta_1$ symmetry.

\begin{figure}[!]
\includegraphics[width=7.5cm,keepaspectratio=true]{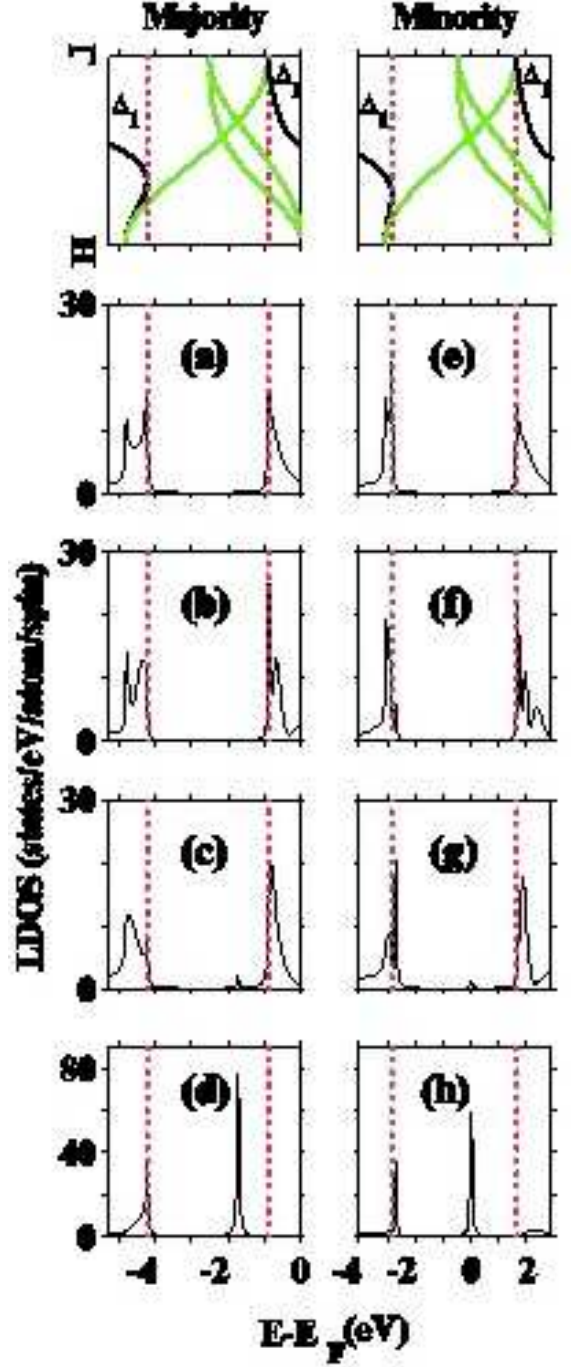}
\caption{(Color online) The top panels represent the band
structures along $\Gamma$-H of bulk Fe for majority and minority
spins. Panels (a-h) give the $\Delta_1$ contribution to the LDOS
$\rho_i(E)$ at $\overline{\Gamma}$ of the Fe$|$vacuum interface as
a function of the layer position. (a) LDOS of an Fe layer at an
infinite distance from the interface; the double peak structure
below -4 eV and single peak (of a double peak structure) above -1
eV correspond to the $\Delta_1$ bulk bands. (b) LDOS of the 6th Fe
layer (the surface layer is layer 1). (c) LDOS of the 3rd Fe
layer; the sharp peak in the bulk band gap corresponds to a
surface state. (d) LDOS of the the surface Fe layer; the surface
state peak has maximum amplitude. (e-h) Same for minority spin. }
\label{fig:delta1}
\end{figure}

\begin{figure}[!]
\includegraphics[width=7.0cm,keepaspectratio=true]{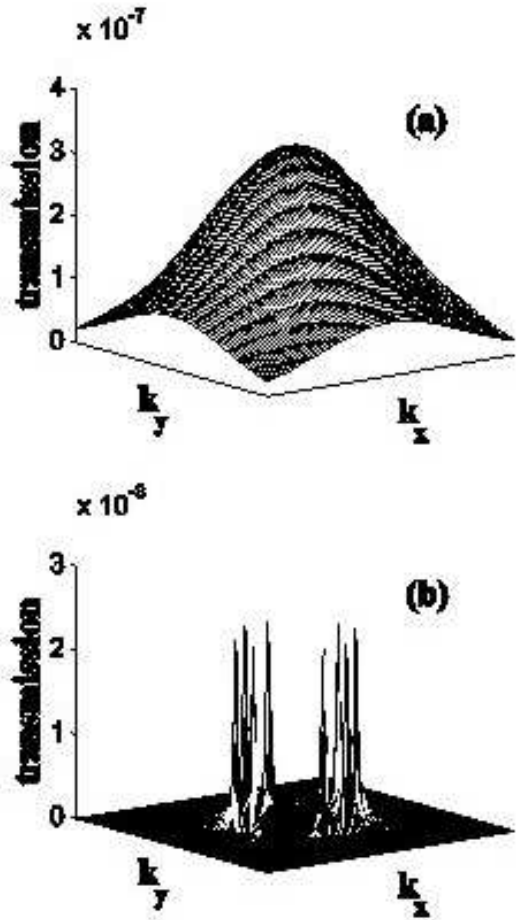}
\caption{Calculated $\mathbf{k}_\parallel$-resolved transmission in (a) the majority and (b)
the minority spin channels through a Fe-vacuum-Fe tunnel junction. The vacuum region has a
thickness of 5.733 \AA\ and the Fe electrodes have a parallel magnetization. Only the central
part of the Brillouin zone is shown, i.e. $-\pi/3a\leq k_x,k_y \leq \pi/3a$; the transmission
in the other parts is close to zero.} \label{fig:transmission}
\end{figure}

This contribution, calculated using the mode matching approach, is shown
in Fig.~\ref{fig:delta1} as a function of the distance from the surface.
In the majority spin LDOS, a sharp peak is found at $E_F-1.8$ eV, which
is very prominent in the surface Fe layer, cf. Fig.~\ref{fig:delta1}(d)
and decays rapidly into the bulk, cf. Figs.~\ref{fig:delta1}(c)-(a).
This evanescent state clearly represents a state that is localized at
the surface. In the minority spin LDOS, a sharp peak with similar
properties is found very close to $E_F$. It corresponds to the surface
state that is observed experimentally by STM.\cite{Stroscio:prl95} The
positions of both these surface state peaks in the majority and minority
LDOS are in excellent agreement with those obtained using a Green
function (KKR) approach.\cite{Uiberacker:prb01,Uiberacker:prb02} Such
surface states are good examples of evanescent states. Clearly, they
are described properly within the mode matching approach which was,
however, disputed in Ref.~\onlinecite{Krstic:prb02}.

Using the mode matching approach, we next calculate the transmission
through an Fe$|$vacuum$|$Fe tunnel junction in which the vacuum region is
only 5.733 \AA~thick (corresponding to two principal layers) and the
magnetizations are parallel. Fig.~\ref{fig:transmission} shows the
calculated $\mathbf{k}_\parallel$-resolved transmission for majority-
and minority-spin channels. The majority conductance has a maximum at
$\overline{\Gamma}$ and it decreases smoothly going away from
$\overline{\Gamma}$. This behavior is quite general for scattering from
a potential barrier. At fixed energy, a wave travelling perpendicular
to the vacuum barrier penetrates furthest into the vacuum and therefore
has a maximum tunneling probability.\cite{Wunnicke:prb02a}

In contrast, the minority conductance is dominated by sharp spikes of very high intensity,
close to 0.13 $\overline{\Gamma}\overline{X}$. This behavior has been analyzed in terms of
interface resonant states that extend relatively far into the vacuum. States originating from
the two Fe$|$vacuum interfaces couple through the thin vacuum region, which enhances the
transmission considerably.\cite{Wunnicke:prb02a} These transmission spikes or ``hot spots''
are observed quite generally in calculations for perfect transition
metal$|$insulator$|$transition metal tunnel junctions.\cite{Butler:prb01,Mathon:prb01} A
calculation using Eq.~\eqref{eq27} gives the same transmission within numerical
accuracy. It again illustrates the correct treatment of evanescent states within the mode
matching approach. We have also tested expression (\ref{eqtt4})
numerically and found that it too gives the same transmission within numerical accuracy.

Finally, the invariance property of the transmission discussed in
Sec.~\ref{invariance}, has been tested numerically by moving the
interfaces between the leads and the scattering region, where the
matching is carried out, into the leads. In this way more and more
slices of lead are treated as scattering region, see
Fig.~\ref{fig:hamiltonian}; the transmission should be invariant under
this operation. Adding up to ten principal layers (20 monolayers) of
bulk Fe to each side of the scattering region changes the transmission
negligibly, by less than one part in $10^8$.

\section{Summary and Conclusions}\label{conclusions}
We have demonstrated the equivalence of the mode matching and the Green
function approaches to calculating the conductance of quantum wires and
interfaces. In the mode matching technique, the scattering problem is
solved by matching the wave function in the scattering region to the
Bloch modes in the leads. The technique is formulated for tight-binding
Hamiltonians and covers all representations that can be expressed in
tight-binding form, including first-principles implementations using
localized orbital basis sets or a real space grid.

Alternatively, the scattering information can be extracted from the
Green function, which is calculated by partitioning the system into a
scattering region and leads. We demonstrate that the Green function
technique can be reformulated in terms of mode matching. In addition we
prove that the mode matching expression for the transmission matrix does
not depend on where in the leads the scattering region is matched to the
ideal leads, which was called into question in Ref.~\onlinecite{Krstic:prb02}.

Calculating the full transmission matrix allows us to study individual scattering amplitudes
from and to every possible mode. We have derived a compact expression for the transmission
matrix elements, Eq.~(\ref{eqtt4}). Only propagating modes enter this expression, since
evanescent modes do not contribute to the transmission directly. The evanescent modes are
important, however, for setting up the correct matching conditions at the boundaries between
the scattering region and the leads. Alternatively, the total transmission probability can be
calculated from the Caroli expression, Eq.~(\ref{eqt4}). Formally this expression sums over
all possible modes, i.e. propagating and evanescent. However, by deriving the Caroli
expression from the mode matching approach we show explicitly that only the propagating modes
give a non-zero contribution.

The mode matching approach and its equivalence to the Green function
approach were illustrated for a simple analytical model, as well as by
numerical calculations on an Fe$|$vacuum$|$Fe tunnel junction. In the
latter we treat the electronic structure within DFT, using a TB-MTO-ASA
basis set. We demonstrate that the layer densities of states that follow
from the mode matching and Green function approaches are numerically
indistinguishable. We identify the Fe(001) surface state in the density
of states and establish its contribution to the transmission through the
tunnel junction. Finally, the invariance property discussed above is
demonstrated numerically on the Fe$|$vacuum$|$Fe tunnel junction.

\appendix
\section{Velocity matrix}
In this appendix we will prove that
\begin{equation}
\mathbf{V}(\pm )_{m,n}=\frac{\hbar }{a}v_{m}(\pm )\delta _{m,n}, \label{eqv2}
\end{equation}
for the expression introduced for the velocity matrices in
Sec.~\ref{transmission}, Eq.~\eqref{eqv1}.
Here, $a$ is the translation period along the wire and, for right-
respectively left-propagating states, $v_{n}(\pm )$ is the Bloch
velocity in the direction of the quantum wire. For evanescent
states $v_{n}(\pm )=0$.

For ease of notation we drop the index $\pm $ in the following.
From Eq.~\eqref{eq6a} and its complex conjugate one has
\begin{eqnarray}
-\mathbf{u}_{m}^{\dagger }\mathbf{Bu}_{n}+\lambda _{n}\mathbf{u}%
_{m}^{\dagger }\left( E\mathbf{I}-\mathbf{H}\right) \mathbf{u}_{n}-(\lambda
_{n})^{2}\mathbf{u}_{m}^{\dagger }\mathbf{B}^{\dagger }\mathbf{u}_{n} &=&0,
\nonumber \\
-\mathbf{u}_{m}^{\dagger }\mathbf{B}^{\dagger }\mathbf{u}_{n}+\lambda _{m}^{\ast
}\mathbf{u}_{m}^{\dagger }\left( E\mathbf{I}-\mathbf{H}\right)
\mathbf{u}_{n}-(\lambda _{m}^{\ast })^{2}\mathbf{u}_{m}^{\dagger }\mathbf{Bu}%
_{n} &=&0. \nonumber \\ \label{eqv3}
\end{eqnarray}
Multiplying the first equation by $\lambda _{m}^{\ast }$, the second by $%
\lambda _{n}$ and subtracting the two gives
\begin{eqnarray}
\left[ \lambda _{n}\mathbf{u}_{m}^{\dagger }\mathbf{B}^{\dagger }\mathbf{u}%
_{n}-\lambda _{m}^{\ast }\mathbf{u}_{m}^{\dagger }\mathbf{Bu}_{n}\right] \left( 1-\lambda
_{m}^{\ast }\lambda _{n}\right) &=&0\Leftrightarrow
\nonumber \\
-i\mathbf{V}_{m,n}\left( 1-\lambda _{m}^{\ast }\lambda _{n}\right) &=&0, \label{eqv4}
\end{eqnarray}
according to Eq.~(\ref{eqv1}). So if $\lambda _{m}^{\ast} \lambda _{n}\neq 1 $ then
$\mathbf{V}_{m,n}=0$.

The velocity matrices contain, by construction, either right- or
left-going modes.
For right-going evanescent modes one has $|\lambda|<1$, for left-going
evanescent modes $|\lambda|>1$, so
$\lambda _{m}^{\ast} \lambda _{n}\neq 1 $ if $n$ and/or $m$ denotes
an evanescent mode; thus $\mathbf{V}_{m,n}=0$.
For propagating states, $\left| \lambda _{m}\right| =1$ and one can
write $\lambda _{m}=\exp (ik_ma)$. It is obvious that if $k_m\neq k_n$,
then $\lambda_{m}^{\ast} \lambda _{n}\neq 1 $ and again
$\mathbf{V}_{m,n}=0$.

This argument does not hold for the diagonal matrix elements of
propagating states, and also not for degenerate propagating states,
i.e. if $\lambda_m = \lambda_n$. In the latter case we take the
derivative $d/dE$ of the first line of Eq.~(\ref{eqv3}).
The coefficients of the terms in $d\mathbf{u}_{n}/dE$ and
$d\mathbf{u}_{m}^{\dagger }/dE$ vanish because $\mathbf{u}_{n}$
and $\mathbf{u}_{m}^{\dagger }$ satisfy the eigenvalue equation
Eq.~(\ref{eq6a}) and its complex conjugate, respectively, and
$\lambda _{n}^{\ast }=1/\lambda _{n} $ for propagating states.
Collecting the remaining terms gives
\begin{eqnarray}
&&-\frac{d\lambda _{n}}{dE}\left[ \lambda _{n}\mathbf{u}_{m}^{\dagger }\mathbf{%
B}^{\dagger }\mathbf{u}_{n}-\lambda _{n}^{\ast }\mathbf{u}_{m}^{\dagger }%
\mathbf{Bu}_{n}\right] +\lambda _{n}\mathbf{u}_{m}^{\dagger }\mathbf{u}_{n}
=0\Leftrightarrow   \nonumber \\
&&i\frac{d\lambda _{n}}{dE}\mathbf{V}_{m,n}+\lambda _{n}\mathbf{u}_{m}^{\dagger
}\mathbf{u}_{n}=0, \label{eqv5}
\end{eqnarray}
where we emphasize that for $m \neq n$ this only holds for degenerate propagating states.
Without loss of generality, degenerate states can be chosen orthogonal, i.e.
$\mathbf{u}_{m}^{\dagger }\mathbf{u}_{n}=0$ for $m\neq n$. Since $d\lambda _{n}/dE\neq 0$, it
then follows that $\mathbf{V}_{m,n}=0$ if $m\neq n$.

The diagonal matrix elements $m=n$ for propagating states are also given by Eq.~(\ref{eqv5}).
Setting $n=m$ and using the fact that the states are normalized, i.e.
$\mathbf{u}_{n}^{\dagger }\mathbf{u}_{n}=1$ gives the expression
\begin{equation}
\mathbf{V}_{n,n}=i\lambda _{n}\frac{dE}{d\lambda _{n}}. \label{eqv5a}
\end{equation}
Since for propagating states we can write $\lambda _{n}=\exp (ika)$ and the Bloch velocity is
defined as $v_{n}=\hbar ^{-1}dE/dk$, this then proves both Eq.~(\ref{eqv2}) and
Eq.~(\ref{eq19a}).

It is straightforward to derive an alternative expression for the velocity matrices in terms
of the advanced matrices
\begin{eqnarray}
\mathbf{V(\pm )}=&&i \left[\mathbf{U}^{a\dagger }(\mp )\mathbf{B}^{\dagger }%
\mathbf{U}^a(\mp )\mathbf{\Lambda }^a(\mp )\right.- \nonumber \\
&&\left. \mathbf{\Lambda }^{a\dagger }(\mp )%
\mathbf{U}^{a\dagger }(\mp )\mathbf{BU}^a(\mp )\right].  \label{eqv6a}
\end{eqnarray}
Multiplying from the left by $\left[\mathbf{\Lambda }^{a\dagger }\right]^{-1}$ and from the
right by $\left[\mathbf{\Lambda }^{a}\right]^{-1}$ this expression is seen to be equivalent
to
\begin{eqnarray}
\mathbf{V(\pm )}=&& i \left[ \{ \mathbf{\Lambda}^{a\dagger}(\mp) \} ^{-1}
\mathbf{U}^{a\dagger}(\mp)\mathbf{B}^{\dagger}\mathbf{U}^a(\mp) \right. - \nonumber \\
&& \left. \mathbf{U}^{a\dagger}(\mp)\mathbf{BU}^a(\mp) \{ \mathbf{\Lambda }^a(\mp) \} ^{-1}
\right] . \label{eqv6}
\end{eqnarray}

\section{Projector matrix}
In order to prove Eq.~(\ref{eqt10b}), we start from Eq.~(\ref{eq14})
and rewrite it using Eq.~(\ref{eqr18a}) as
\begin{equation}
\mathbf{Q}_{0} =\mathbf{B}_\mathrm{L}\mathbf{F}_\mathrm{L}^{-1}(+)
-\left[\mathbf{F}_\mathrm{L}^{a\dagger}(-)\right]^{-1}\mathbf{B}_\mathrm{L}^\dagger.
\label{eqb1}
\end{equation}
In the following we will drop the subscript L for ease of notation.
Multiplying \eqref{eqb1} on the left with the identity operator
$\mathbf{I}=\sum \mathbf{u}_n(+)\widetilde{\mathbf{u}}_n^\dagger(+)$
and on the right with
$\mathbf{I}=\sum \widetilde{\mathbf{u}}_m^{a}(-)\mathbf{u}_m^{a\dagger}(-)$
yields
\begin{eqnarray}
&&\mathbf{Q}_{0} = \sum_{m,n=1}^N \widetilde{\mathbf{u}}_m^a(-)
\widetilde{\mathbf{u}}_n^\dagger(+) \label{eqb2} \\
&&\left[\frac{1}{\lambda_n(+)}\mathbf{u}_m^{a\dagger}(-) \mathbf{B}\mathbf{u}_n(+) -
\frac{1}{\lambda_m^{a*}(-)}\mathbf{u}_m^{a\dagger}(-)
\mathbf{B}^\dagger\mathbf{u}_n(+)\right], \nonumber
\end{eqnarray}
where use has been made of Eqs.~\eqref{eq8}, \eqref{eq7} and \eqref{eqr13a}.

By similar arguments as those leading to Eq.~(\ref{eqv4}) it follows that $[\ldots]=0$,
unless $\lambda_m^a(-)=1/\lambda_n^*(+)$. From Eq.~(\ref{eqr13a}) it follows that this is
true if $n=m$ and $n$ denotes a propagating mode. In addition, for every $(+)$ evanescent
mode $n$ with eigenvalue $\lambda_n(+)$, there is one $(-)$ evanescent mode with eigenvalue
$\lambda_n(-)=1/\lambda_n^*(+)$.\cite{Molinari:jpa97} Again it follows from
Eq.~(\ref{eqr13a}) that only the $m=n$ terms are non-zero for evanescent modes in
Eq.~(\ref{eqb2}). This equation then simplifies to
\begin{eqnarray}
&&\mathbf{Q}_{0} = \sum_{n=1}^{N_p} \widetilde{\mathbf{u}}^a_n(-)
\widetilde{\mathbf{u}}_n^\dagger(+) \label{eqb3} \\
&&\left[\frac{1}{\lambda_n(+)}\mathbf{u}_n^\dagger(+) \mathbf{B}\mathbf{u}_n(+) -
\lambda_n(+)\mathbf{u}_n^\dagger(+) \mathbf{B}^\dagger\mathbf{u}_n(+)\right] \nonumber \\
&&+ \sum_{n=N_p+1}^N \widetilde{\mathbf{u}}^a_n(-) \widetilde{\mathbf{u}}_n^\dagger(+) \nonumber \\
&&\left[\frac{1}{\lambda_n(+)}\mathbf{u}_n^\dagger(-) \mathbf{B}\mathbf{u}_n(+) -
\lambda_n(+)\mathbf{u}_n^\dagger(-) \mathbf{B}^\dagger\mathbf{u}_n(+)\right], \nonumber
\end{eqnarray}
where the first summation is over the $N_p$ propagating modes and the second summation is
over the $N-N_p$ evanescent modes.

The projector matrix $\mathbf{P}$ is defined by Eq.~(\ref{eqt10a}). In the product
$\mathbf{Q}_0\mathbf{P}$ only the first summation in Eq.~(\ref{eqb3}) survives. Moreover,
since $[\ldots]=i\mathbf{V}_{n,n}$ according to Eq.~(\ref{eqv4}), we have
\begin{equation}
\mathbf{Q}_{0}\mathbf{P} = i \sum_{n=1}^{N_p} \widetilde{\mathbf{u}}^a_n(-)
\widetilde{\mathbf{u}}_n^{a\dagger}(-)\mathbf{V}_{n,n}(+). \label{eqb4}
\end{equation}
Together with Eq.~(\ref{eqt9a}) this then proves Eq.~(\ref{eqt10b}).

\acknowledgments

This work was financially supported by the ``Nederlandse Organisatie voor Wetenschappelijk
Onderzoek (NWO)'' via the research programs of ``Chemische Wetenschappen (CW)'' and the
``Stichting voor Fundamenteel Onderzoek der Materie (FOM)''; by ``NanoNed'', a nanotechnology
programme of the Dutch Ministry of Economic Affairs; by the European Commission's Research
Training Network ``Computational Magnetoelectronics'' (contract No. HPRN-CT-2000-00143); and
by the NEDO International Joint Research program ``Nano-scale Magnetoelectronics''. Part of
the calculations were performed with a grant of computer time from the ``Stichting Nationale
Computerfaciliteiten (NCF)''.

%\bibliography{pjk}

\begin{thebibliography}{50}
\expandafter\ifx\csname natexlab\endcsname\relax\def\natexlab#1{#1}\fi
\expandafter\ifx\csname bibnamefont\endcsname\relax
  \def\bibnamefont#1{#1}\fi
\expandafter\ifx\csname bibfnamefont\endcsname\relax
  \def\bibfnamefont#1{#1}\fi
\expandafter\ifx\csname citenamefont\endcsname\relax
  \def\citenamefont#1{#1}\fi
\expandafter\ifx\csname url\endcsname\relax
  \def\url#1{\texttt{#1}}\fi
\expandafter\ifx\csname urlprefix\endcsname\relax\def\urlprefix{URL }\fi
\providecommand{\bibinfo}[2]{#2} \providecommand{\eprint}[2][]{\url{#2}}

\bibitem[{\citenamefont{Baibich et~al.}(1988)\citenamefont{Baibich, Broto,
  Fert, van Dau, Petroff, Etienne, Creuzet, Friedrich, and
  Chazelas}}]{Baibich:prl88}
\bibinfo{author}{\bibfnamefont{M.~N.} \bibnamefont{Baibich}},
  \bibinfo{author}{\bibfnamefont{J.~M.} \bibnamefont{Broto}},
  \bibinfo{author}{\bibfnamefont{A.}~\bibnamefont{Fert}},
  \bibinfo{author}{\bibfnamefont{F.~N.} \bibnamefont{van Dau}},
  \bibinfo{author}{\bibfnamefont{F.}~\bibnamefont{Petroff}},
  \bibinfo{author}{\bibfnamefont{P.}~\bibnamefont{Etienne}},
  \bibinfo{author}{\bibfnamefont{G.}~\bibnamefont{Creuzet}},
  \bibinfo{author}{\bibfnamefont{A.}~\bibnamefont{Friedrich}},
  \bibnamefont{and} \bibinfo{author}{\bibfnamefont{J.}~\bibnamefont{Chazelas}},
  \bibinfo{journal}{Phys. Rev. Lett.} \textbf{\bibinfo{volume}{61}},
  \bibinfo{pages}{2472} (\bibinfo{year}{1988}).

\bibitem[{\citenamefont{Binasch et~al.}(1989)\citenamefont{Binasch,
  Gr{\"{u}}nberg, Saurenbach, and Zinn}}]{Binasch:prb89}
\bibinfo{author}{\bibfnamefont{G.}~\bibnamefont{Binasch}},
  \bibinfo{author}{\bibfnamefont{P.}~\bibnamefont{Gr{\"{u}}nberg}},
  \bibinfo{author}{\bibfnamefont{F.}~\bibnamefont{Saurenbach}},
  \bibnamefont{and} \bibinfo{author}{\bibfnamefont{W.}~\bibnamefont{Zinn}},
  \bibinfo{journal}{Phys. Rev. B} \textbf{\bibinfo{volume}{39}},
  \bibinfo{pages}{4828} (\bibinfo{year}{1989}).

\bibitem[{\citenamefont{Agra{\"{i}}t et~al.}(2003)\citenamefont{Agra{\"{i}}t,
  Yeyati, and van Ruitenbeek}}]{Agrait:prp03}
\bibinfo{author}{\bibfnamefont{N.}~\bibnamefont{Agra{\"{i}}t}},
  \bibinfo{author}{\bibfnamefont{A.~L.} \bibnamefont{Yeyati}},
  \bibnamefont{and} \bibinfo{author}{\bibfnamefont{J.~M.} \bibnamefont{van
  Ruitenbeek}}, \bibinfo{journal}{Phys. Rep.} \textbf{\bibinfo{volume}{377}},
  \bibinfo{pages}{81} (\bibinfo{year}{2003}).

\bibitem[{\citenamefont{B{\"{u}}ttiker
  et~al.}(1985)\citenamefont{B{\"{u}}ttiker, Imry, Landauer, and
  Pinhas}}]{Buttiker:prb85}
\bibinfo{author}{\bibfnamefont{M.}~\bibnamefont{B{\"{u}}ttiker}},
  \bibinfo{author}{\bibfnamefont{Y.}~\bibnamefont{Imry}},
  \bibinfo{author}{\bibfnamefont{R.}~\bibnamefont{Landauer}}, \bibnamefont{and}
  \bibinfo{author}{\bibfnamefont{S.}~\bibnamefont{Pinhas}},
  \bibinfo{journal}{Phys. Rev. B} \textbf{\bibinfo{volume}{31}},
  \bibinfo{pages}{6207} (\bibinfo{year}{1985}).

\bibitem[{\citenamefont{Schep et~al.}(1997)\citenamefont{Schep, van Hoof,
  Kelly, Bauer, and Inglesfield}}]{Schep:prb97}
\bibinfo{author}{\bibfnamefont{K.~M.} \bibnamefont{Schep}},
  \bibinfo{author}{\bibfnamefont{J.~B. A.~N.} \bibnamefont{van Hoof}},
  \bibinfo{author}{\bibfnamefont{P.~J.} \bibnamefont{Kelly}},
  \bibinfo{author}{\bibfnamefont{G.~E.~W.} \bibnamefont{Bauer}},
  \bibnamefont{and} \bibinfo{author}{\bibfnamefont{J.~E.}
  \bibnamefont{Inglesfield}}, \bibinfo{journal}{Phys. Rev. B}
  \textbf{\bibinfo{volume}{56}}, \bibinfo{pages}{10805} (\bibinfo{year}{1997}).

\bibitem[{\citenamefont{van Hoof et~al.}(1999)\citenamefont{van Hoof, Schep,
  Brataas, Bauer, and Kelly}}]{vanHoof:prb99}
\bibinfo{author}{\bibfnamefont{J.~B. A.~N.} \bibnamefont{van Hoof}},
  \bibinfo{author}{\bibfnamefont{K.~M.} \bibnamefont{Schep}},
  \bibinfo{author}{\bibfnamefont{A.}~\bibnamefont{Brataas}},
  \bibinfo{author}{\bibfnamefont{G.~E.~W.} \bibnamefont{Bauer}},
  \bibnamefont{and} \bibinfo{author}{\bibfnamefont{P.~J.} \bibnamefont{Kelly}},
  \bibinfo{journal}{Phys. Rev. B} \textbf{\bibinfo{volume}{59}},
  \bibinfo{pages}{138} (\bibinfo{year}{1999}).

\bibitem[{\citenamefont{Kudrnovsk\'{y}
  et~al.}(2000)\citenamefont{Kudrnovsk\'{y}, Drchal, Blaas, Weinberger, Turek,
  and Bruno}}]{Kudrnovsky:prb00}
\bibinfo{author}{\bibfnamefont{J.}~\bibnamefont{Kudrnovsk\'{y}}},
  \bibinfo{author}{\bibfnamefont{V.}~\bibnamefont{Drchal}},
  \bibinfo{author}{\bibfnamefont{C.}~\bibnamefont{Blaas}},
  \bibinfo{author}{\bibfnamefont{P.}~\bibnamefont{Weinberger}},
  \bibinfo{author}{\bibfnamefont{I.}~\bibnamefont{Turek}}, \bibnamefont{and}
  \bibinfo{author}{\bibfnamefont{P.}~\bibnamefont{Bruno}},
  \bibinfo{journal}{Phys. Rev. B} \textbf{\bibinfo{volume}{62}},
  \bibinfo{pages}{15084} (\bibinfo{year}{2000}).

\bibitem[{\citenamefont{Xia et~al.}(2001)\citenamefont{Xia, Kelly, Bauer,
  Turek, Kudrnovsk\'{y}, and Drchal}}]{Xia:prb01}
\bibinfo{author}{\bibfnamefont{K.}~\bibnamefont{Xia}},
  \bibinfo{author}{\bibfnamefont{P.~J.} \bibnamefont{Kelly}},
  \bibinfo{author}{\bibfnamefont{G.}~\bibnamefont{Bauer}},
  \bibinfo{author}{\bibfnamefont{I.}~\bibnamefont{Turek}},
  \bibinfo{author}{\bibfnamefont{J.}~\bibnamefont{Kudrnovsk\'{y}}},
  \bibnamefont{and} \bibinfo{author}{\bibfnamefont{V.}~\bibnamefont{Drchal}},
  \bibinfo{journal}{Phys. Rev. B} \textbf{\bibinfo{volume}{63}},
  \bibinfo{pages}{064407} (\bibinfo{year}{2001}).

\bibitem[{\citenamefont{Riedel et~al.}(2001)\citenamefont{Riedel, Zahn, and
  Mertig}}]{Riedel:prb01}
\bibinfo{author}{\bibfnamefont{I.}~\bibnamefont{Riedel}},
  \bibinfo{author}{\bibfnamefont{P.}~\bibnamefont{Zahn}}, \bibnamefont{and}
  \bibinfo{author}{\bibfnamefont{I.}~\bibnamefont{Mertig}},
  \bibinfo{journal}{Phys. Rev. B} \textbf{\bibinfo{volume}{63}},
  \bibinfo{pages}{195403} (\bibinfo{year}{2001}).

\bibitem[{\citenamefont{Taylor et~al.}(2001)\citenamefont{Taylor, Guo, and
  Wang}}]{Taylor:prb01}
\bibinfo{author}{\bibfnamefont{J.}~\bibnamefont{Taylor}},
  \bibinfo{author}{\bibfnamefont{H.}~\bibnamefont{Guo}}, \bibnamefont{and}
  \bibinfo{author}{\bibfnamefont{J.}~\bibnamefont{Wang}},
  \bibinfo{journal}{Phys. Rev. B} \textbf{\bibinfo{volume}{63}},
  \bibinfo{pages}{245407} (\bibinfo{year}{2001}).

\bibitem[{\citenamefont{Brandbyge et~al.}(2002)\citenamefont{Brandbyge, Mozos,
  Ordej\'{o}n, Taylor, and Stokbro}}]{Brandbyge:prb02}
\bibinfo{author}{\bibfnamefont{M.}~\bibnamefont{Brandbyge}},
  \bibinfo{author}{\bibfnamefont{J.~L.} \bibnamefont{Mozos}},
  \bibinfo{author}{\bibfnamefont{P.}~\bibnamefont{Ordej\'{o}n}},
  \bibinfo{author}{\bibfnamefont{J.}~\bibnamefont{Taylor}}, \bibnamefont{and}
  \bibinfo{author}{\bibfnamefont{K.}~\bibnamefont{Stokbro}},
  \bibinfo{journal}{Phys. Rev. B} \textbf{\bibinfo{volume}{65}},
  \bibinfo{pages}{165401} (\bibinfo{year}{2002}).

\bibitem[{\citenamefont{Wortmann et~al.}(2002)\citenamefont{Wortmann, Ishida,
  and Bl{\"{u}}gel}}]{Wortmann:prb02b}
\bibinfo{author}{\bibfnamefont{D.}~\bibnamefont{Wortmann}},
  \bibinfo{author}{\bibfnamefont{H.}~\bibnamefont{Ishida}}, \bibnamefont{and}
  \bibinfo{author}{\bibfnamefont{S.}~\bibnamefont{Bl{\"{u}}gel}},
  \bibinfo{journal}{Phys. Rev. B} \textbf{\bibinfo{volume}{66}},
  \bibinfo{pages}{075113} (\bibinfo{year}{2002}).

\bibitem[{\citenamefont{Thygesen et~al.}(2003)\citenamefont{Thygesen,
  Bollinger, and Jacobsen}}]{Thygesen:prb03}
\bibinfo{author}{\bibfnamefont{K.~S.} \bibnamefont{Thygesen}},
  \bibinfo{author}{\bibfnamefont{M.~V.} \bibnamefont{Bollinger}},
  \bibnamefont{and} \bibinfo{author}{\bibfnamefont{K.~W.}
  \bibnamefont{Jacobsen}}, \bibinfo{journal}{Phys. Rev. B}
  \textbf{\bibinfo{volume}{67}}, \bibinfo{pages}{115404}
  (\bibinfo{year}{2003}).

\bibitem[{\citenamefont{Mavropoulos et~al.}(2004)\citenamefont{Mavropoulos,
  Papanikolaou, and Dederichs}}]{Mavropoulos:prb04}
\bibinfo{author}{\bibfnamefont{P.}~\bibnamefont{Mavropoulos}},
  \bibinfo{author}{\bibfnamefont{N.}~\bibnamefont{Papanikolaou}},
  \bibnamefont{and}
  \bibinfo{author}{\bibfnamefont{P.}~\bibnamefont{Dederichs}},
  \bibinfo{journal}{Phys. Rev. B} \textbf{\bibinfo{volume}{69}},
  \bibinfo{pages}{125104} (\bibinfo{year}{2004}).

\bibitem[{\citenamefont{Caroli et~al.}(1971)\citenamefont{Caroli, Combescot,
  Nozi\`{e}res, and Saint-James}}]{Caroli:jpc71}
\bibinfo{author}{\bibfnamefont{C.}~\bibnamefont{Caroli}},
  \bibinfo{author}{\bibfnamefont{R.}~\bibnamefont{Combescot}},
  \bibinfo{author}{\bibfnamefont{P.}~\bibnamefont{Nozi\`{e}res}},
  \bibnamefont{and}
  \bibinfo{author}{\bibfnamefont{D.}~\bibnamefont{Saint-James}},
  \bibinfo{journal}{J. Phys. C: Sol. State Phys.} \textbf{\bibinfo{volume}{4}},
  \bibinfo{pages}{916} (\bibinfo{year}{1971}).

\bibitem[{\citenamefont{Datta}(1995)}]{Datta:95}
\bibinfo{author}{\bibfnamefont{S.}~\bibnamefont{Datta}},
  \emph{\bibinfo{title}{Electronic Transport in Mesoscopic Systems}}
  (\bibinfo{publisher}{Cambridge University Press},
  \bibinfo{address}{Cambridge}, \bibinfo{year}{1995}).

\bibitem[{\citenamefont{Ando}(1991)}]{Ando:prb91}
\bibinfo{author}{\bibfnamefont{T.}~\bibnamefont{Ando}}, \bibinfo{journal}{Phys.
  Rev. B} \textbf{\bibinfo{volume}{44}}, \bibinfo{pages}{8017}
  (\bibinfo{year}{1991}).

\bibitem[{\citenamefont{Nicoli{\'c} and MacKinnon}(1994)}]{Nicolic:prb94}
\bibinfo{author}{\bibfnamefont{K.}~\bibnamefont{Nicoli{\'c}}} \bibnamefont{and}
  \bibinfo{author}{\bibfnamefont{A.}~\bibnamefont{MacKinnon}},
  \bibinfo{journal}{Phys. Rev. B} \textbf{\bibinfo{volume}{50}},
  \bibinfo{pages}{11008} (\bibinfo{year}{1994}).

\bibitem[{\citenamefont{Xia et~al.}(2002{\natexlab{a}})\citenamefont{Xia,
  Kelly, Bauer, Brataas, and Turek}}]{Xia:prb02}
\bibinfo{author}{\bibfnamefont{K.}~\bibnamefont{Xia}},
  \bibinfo{author}{\bibfnamefont{P.~J.} \bibnamefont{Kelly}},
  \bibinfo{author}{\bibfnamefont{G.~E.~W.} \bibnamefont{Bauer}},
  \bibinfo{author}{\bibfnamefont{A.}~\bibnamefont{Brataas}}, \bibnamefont{and}
  \bibinfo{author}{\bibfnamefont{I.}~\bibnamefont{Turek}},
  \bibinfo{journal}{Phys. Rev. B} \textbf{\bibinfo{volume}{65}},
  \bibinfo{pages}{220401} (\bibinfo{year}{2002}{\natexlab{a}}).

\bibitem[{\citenamefont{Xia et~al.}(2002{\natexlab{b}})\citenamefont{Xia,
  Kelly, Bauer, and Turek}}]{Xia:prl02}
\bibinfo{author}{\bibfnamefont{K.}~\bibnamefont{Xia}},
  \bibinfo{author}{\bibfnamefont{P.~J.} \bibnamefont{Kelly}},
  \bibinfo{author}{\bibfnamefont{G.~E.~W.} \bibnamefont{Bauer}},
  \bibnamefont{and} \bibinfo{author}{\bibfnamefont{I.}~\bibnamefont{Turek}},
  \bibinfo{journal}{Phys. Rev. Lett.} \textbf{\bibinfo{volume}{89}},
  \bibinfo{pages}{166603} (\bibinfo{year}{2002}{\natexlab{b}}).

\bibitem[{\citenamefont{Zwierzycki et~al.}(2003)\citenamefont{Zwierzycki, Xia,
  Kelly, and Bauer}}]{Zwierzycki:prb03}
\bibinfo{author}{\bibfnamefont{M.}~\bibnamefont{Zwierzycki}},
  \bibinfo{author}{\bibfnamefont{K.}~\bibnamefont{Xia}},
  \bibinfo{author}{\bibfnamefont{P.~J.} \bibnamefont{Kelly}}, \bibnamefont{and}
  \bibinfo{author}{\bibfnamefont{G.~E.~W.} \bibnamefont{Bauer}},
  \bibinfo{journal}{Phys. Rev. B} \textbf{\bibinfo{volume}{67}},
  \bibinfo{pages}{092401} (\bibinfo{year}{2003}).

\bibitem[{\citenamefont{Krsti{\'{c}} et~al.}(2002)\citenamefont{Krsti{\'{c}},
  Zhang, and Butler}}]{Krstic:prb02}
\bibinfo{author}{\bibfnamefont{P.~S.} \bibnamefont{Krsti{\'{c}}}},
  \bibinfo{author}{\bibfnamefont{X.-G.} \bibnamefont{Zhang}}, \bibnamefont{and}
  \bibinfo{author}{\bibfnamefont{W.~H.} \bibnamefont{Butler}},
  \bibinfo{journal}{Phys. Rev. B} \textbf{\bibinfo{volume}{66}},
  \bibinfo{pages}{205319} (\bibinfo{year}{2002}).

\bibitem[{\citenamefont{Sautet and Joachim}(1988)}]{Sautet:prb88}
\bibinfo{author}{\bibfnamefont{P.}~\bibnamefont{Sautet}} \bibnamefont{and}
  \bibinfo{author}{\bibfnamefont{C.}~\bibnamefont{Joachim}},
  \bibinfo{journal}{Phys. Rev. B} \textbf{\bibinfo{volume}{38}},
  \bibinfo{pages}{12238} (\bibinfo{year}{1988}).

\bibitem[{Not({\natexlab{a}})}]{Note:ortho_remark}
\bibinfo{note}{We restrict ourselves to a representation on orthogonal basis
  sets, but the extension to non-orthogonal basis sets is straightforward}.

\bibitem[{\citenamefont{Xia et~al.}()\citenamefont{Xia, Zwierzycki, Talanana,
  Kelly, and Bauer}}]{Xia:prb05}
\bibinfo{author}{\bibfnamefont{K.}~\bibnamefont{Xia}},
  \bibinfo{author}{\bibfnamefont{M.}~\bibnamefont{Zwierzycki}},
  \bibinfo{author}{\bibfnamefont{M.}~\bibnamefont{Talanana}},
  \bibinfo{author}{\bibfnamefont{P.~J.} \bibnamefont{Kelly}}, \bibnamefont{and}
  \bibinfo{author}{\bibfnamefont{G.~E.~W.} \bibnamefont{Bauer}},
  \bibinfo{note}{to be published}.

\bibitem[{\citenamefont{Khomyakov and Brocks}(2004)}]{Khomyakov:prb04}
\bibinfo{author}{\bibfnamefont{P.~A.} \bibnamefont{Khomyakov}}
  \bibnamefont{and} \bibinfo{author}{\bibfnamefont{G.}~\bibnamefont{Brocks}},
  \bibinfo{journal}{Phys. Rev. B} \textbf{\bibinfo{volume}{70}},
  \bibinfo{pages}{195402} (\bibinfo{year}{2004}).

\bibitem[{\citenamefont{MacKinnon}(1985)}]{MacKinnon:zfp85}
\bibinfo{author}{\bibfnamefont{A.}~\bibnamefont{MacKinnon}},
  \bibinfo{journal}{Z. Phys. B} \textbf{\bibinfo{volume}{59}},
  \bibinfo{pages}{385} (\bibinfo{year}{1985}).

\bibitem[{Not({\natexlab{b}})}]{Note:singular_remark}
\bibinfo{note}{In some representations the hopping matrices can be singular. In
  all cases where the inverse of such a matrix would be needed, one can use the
  well-defined pseudo-inverse matrix instead, cf. Ref.~\onlinecite{Golub:96}}.

\bibitem[{\citenamefont{Tisseur and Meerbergen}(2001)}]{Tisseur:siam01}
\bibinfo{author}{\bibfnamefont{F.}~\bibnamefont{Tisseur}} \bibnamefont{and}
  \bibinfo{author}{\bibfnamefont{K.}~\bibnamefont{Meerbergen}},
  \bibinfo{journal}{SIAM Review} \textbf{\bibinfo{volume}{43}},
  \bibinfo{pages}{235} (\bibinfo{year}{2001}).

\bibitem[{\citenamefont{Golub and van Loan}(1996)}]{Golub:96}
\bibinfo{author}{\bibfnamefont{G.}~\bibnamefont{Golub}} \bibnamefont{and}
  \bibinfo{author}{\bibfnamefont{C.~F.} \bibnamefont{van Loan}},
  \emph{\bibinfo{title}{Matrix Computations}} (\bibinfo{publisher}{Johns
  Hopkins University Press}, \bibinfo{address}{Baltimore},
  \bibinfo{year}{1996}).

\bibitem[{\citenamefont{Molinari}(1997)}]{Molinari:jpa97}
\bibinfo{author}{\bibfnamefont{L.}~\bibnamefont{Molinari}},
  \bibinfo{journal}{J. Phys. A: Math. Gen.} \textbf{\bibinfo{volume}{30}},
  \bibinfo{pages}{983} (\bibinfo{year}{1997}).

\bibitem[{Not({\natexlab{c}})}]{Note:current_remark}
\bibinfo{note}{The current per mode is given by velocity $\times$ density.
  Since we normalize a mode in a slice, its density is given by 1/thickness of
  the slice.}

\bibitem[{\citenamefont{Godfrin}(1991)}]{Godfrin:jp91}
\bibinfo{author}{\bibfnamefont{E.}~\bibnamefont{Godfrin}}, \bibinfo{journal}{J.
  Phys.: Condens. Matter.} \textbf{\bibinfo{volume}{3}}, \bibinfo{pages}{7843}
  (\bibinfo{year}{1991}).

\bibitem[{\citenamefont{Williams et~al.}(1982)\citenamefont{Williams,
  Feibelman, and Lang}}]{Williams:prb82}
\bibinfo{author}{\bibfnamefont{A.~R.} \bibnamefont{Williams}},
  \bibinfo{author}{\bibfnamefont{P.~J.} \bibnamefont{Feibelman}},
  \bibnamefont{and} \bibinfo{author}{\bibfnamefont{N.~D.} \bibnamefont{Lang}},
  \bibinfo{journal}{Phys. Rev. B} \textbf{\bibinfo{volume}{26}},
  \bibinfo{pages}{5433} (\bibinfo{year}{1982}).

\bibitem[{\citenamefont{Turek et~al.}(1997)\citenamefont{Turek, Drchal,
  Kudrnovsk\'{y}, \v{S}ob, and Weinberger}}]{Turek:97}
\bibinfo{author}{\bibfnamefont{I.}~\bibnamefont{Turek}},
  \bibinfo{author}{\bibfnamefont{V.}~\bibnamefont{Drchal}},
  \bibinfo{author}{\bibfnamefont{J.}~\bibnamefont{Kudrnovsk\'{y}}},
  \bibinfo{author}{\bibfnamefont{M.}~\bibnamefont{\v{S}ob}}, \bibnamefont{and}
  \bibinfo{author}{\bibfnamefont{P.}~\bibnamefont{Weinberger}},
  \emph{\bibinfo{title}{Electronic Structure of Disordered Alloys, Surfaces and
  Interfaces}} (\bibinfo{publisher}{Kluwer},
  \bibinfo{address}{Boston-London-Dordrecht}, \bibinfo{year}{1997}).

\bibitem[{\citenamefont{Guinea et~al.}(1983)\citenamefont{Guinea, Tejedor,
  Flores, and Louis}}]{Guinea:prb83}
\bibinfo{author}{\bibfnamefont{F.}~\bibnamefont{Guinea}},
  \bibinfo{author}{\bibfnamefont{C.}~\bibnamefont{Tejedor}},
  \bibinfo{author}{\bibfnamefont{F.}~\bibnamefont{Flores}}, \bibnamefont{and}
  \bibinfo{author}{\bibfnamefont{E.}~\bibnamefont{Louis}},
  \bibinfo{journal}{Phys. Rev. B} \textbf{\bibinfo{volume}{28}},
  \bibinfo{pages}{4397} (\bibinfo{year}{1983}).

\bibitem[{\citenamefont{Fisher and Lee}(1981)}]{Fisher:prb81}
\bibinfo{author}{\bibfnamefont{D.~S.} \bibnamefont{Fisher}} \bibnamefont{and}
  \bibinfo{author}{\bibfnamefont{P.~A.} \bibnamefont{Lee}},
  \bibinfo{journal}{Phys. Rev. B} \textbf{\bibinfo{volume}{23}},
  \bibinfo{pages}{6851} (\bibinfo{year}{1981}).

\bibitem[{\citenamefont{Messiah}(1961)}]{Messiah:61}
\bibinfo{author}{\bibfnamefont{A.}~\bibnamefont{Messiah}},
  \emph{\bibinfo{title}{Quantum Mechanics}}
  (\bibinfo{publisher}{North-Holland}, \bibinfo{address}{Amsterdam},
  \bibinfo{year}{1961}).

\bibitem[{\citenamefont{Khomyakov}()}]{Khomyakov:unpub}
\bibinfo{author}{\bibfnamefont{P.~A.} \bibnamefont{Khomyakov}},
  \bibinfo{note}{(unpublished)}.

\bibitem[{\citenamefont{Kostyrko}(2000)}]{Kostyrko:prb00}
\bibinfo{author}{\bibfnamefont{T.}~\bibnamefont{Kostyrko}},
  \bibinfo{journal}{Phys. Rev. B} \textbf{\bibinfo{volume}{62}},
  \bibinfo{pages}{2458} (\bibinfo{year}{2000}).

\bibitem[{\citenamefont{Landau and Lifschitz}(1981)}]{Landau:81}
\bibinfo{author}{\bibfnamefont{L.~D.} \bibnamefont{Landau}} \bibnamefont{and}
  \bibinfo{author}{\bibfnamefont{E.~M.} \bibnamefont{Lifschitz}},
  \emph{\bibinfo{title}{Quantum Mechanics: Non-relativistic Theory}}
  (\bibinfo{publisher}{Pergamon Press}, \bibinfo{address}{Oxford},
  \bibinfo{year}{1981}), \bibinfo{note}{\S 128}.

\bibitem[{\citenamefont{Perdew and Zunger}(1981)}]{Perdew:prb81}
\bibinfo{author}{\bibfnamefont{J.~P.} \bibnamefont{Perdew}} \bibnamefont{and}
  \bibinfo{author}{\bibfnamefont{A.}~\bibnamefont{Zunger}},
  \bibinfo{journal}{Phys. Rev. B} \textbf{\bibinfo{volume}{23}},
  \bibinfo{pages}{5048} (\bibinfo{year}{1981}).

\bibitem[{\citenamefont{Andersen et~al.}(1985)\citenamefont{Andersen, Jepsen,
  and Gl{\"{o}}tzel}}]{Andersen:85}
\bibinfo{author}{\bibfnamefont{O.~K.} \bibnamefont{Andersen}},
  \bibinfo{author}{\bibfnamefont{O.}~\bibnamefont{Jepsen}}, \bibnamefont{and}
  \bibinfo{author}{\bibfnamefont{D.}~\bibnamefont{Gl{\"{o}}tzel}}, in
  \emph{\bibinfo{booktitle}{Highlights of Condensed Matter Theory}}, edited by
  \bibinfo{editor}{\bibfnamefont{F.}~\bibnamefont{Bassani}},
  \bibinfo{editor}{\bibfnamefont{F.}~\bibnamefont{Fumi}}, \bibnamefont{and}
  \bibinfo{editor}{\bibfnamefont{M.~P.} \bibnamefont{Tosi}}
  (\bibinfo{publisher}{North-Holland}, \bibinfo{address}{Amsterdam},
  \bibinfo{year}{1985}), pp. \bibinfo{pages}{59--176}.

\bibitem[{\citenamefont{Wang and Freeman}(1981)}]{Wang:prb81}
\bibinfo{author}{\bibfnamefont{C.~S.} \bibnamefont{Wang}} \bibnamefont{and}
  \bibinfo{author}{\bibfnamefont{A.~J.} \bibnamefont{Freeman}},
  \bibinfo{journal}{Phys. Rev. B} \textbf{\bibinfo{volume}{24}},
  \bibinfo{pages}{4364} (\bibinfo{year}{1981}).

\bibitem[{\citenamefont{Stroscio et~al.}(1995)\citenamefont{Stroscio, Pierce,
  Davies, Celotta, and Weinert}}]{Stroscio:prl95}
\bibinfo{author}{\bibfnamefont{J.~A.} \bibnamefont{Stroscio}},
  \bibinfo{author}{\bibfnamefont{D.~T.} \bibnamefont{Pierce}},
  \bibinfo{author}{\bibfnamefont{A.}~\bibnamefont{Davies}},
  \bibinfo{author}{\bibfnamefont{R.~J.} \bibnamefont{Celotta}},
  \bibnamefont{and} \bibinfo{author}{\bibfnamefont{M.}~\bibnamefont{Weinert}},
  \bibinfo{journal}{Phys. Rev. Lett.} \textbf{\bibinfo{volume}{75}},
  \bibinfo{pages}{2960} (\bibinfo{year}{1995}).

\bibitem[{\citenamefont{Uiberacker and Levy}(2001)}]{Uiberacker:prb01}
\bibinfo{author}{\bibfnamefont{C.}~\bibnamefont{Uiberacker}} \bibnamefont{and}
  \bibinfo{author}{\bibfnamefont{P.~M.} \bibnamefont{Levy}},
  \bibinfo{journal}{Phys. Rev. B} \textbf{\bibinfo{volume}{64}},
  \bibinfo{pages}{193404} (\bibinfo{year}{2001}).

\bibitem[{\citenamefont{Uiberacker and Levy}(2002)}]{Uiberacker:prb02}
\bibinfo{author}{\bibfnamefont{C.}~\bibnamefont{Uiberacker}} \bibnamefont{and}
  \bibinfo{author}{\bibfnamefont{P.~M.} \bibnamefont{Levy}},
  \bibinfo{journal}{Phys. Rev. B} \textbf{\bibinfo{volume}{65}},
  \bibinfo{pages}{169904(E)} (\bibinfo{year}{2002}).

\bibitem[{\citenamefont{Wunnicke et~al.}(2002)\citenamefont{Wunnicke,
  Papanikolaou, Zeller, Dederichs, Drchal, and
  Kudrnovsk\'{y}}}]{Wunnicke:prb02a}
\bibinfo{author}{\bibfnamefont{O.}~\bibnamefont{Wunnicke}},
  \bibinfo{author}{\bibfnamefont{N.}~\bibnamefont{Papanikolaou}},
  \bibinfo{author}{\bibfnamefont{R.}~\bibnamefont{Zeller}},
  \bibinfo{author}{\bibfnamefont{P.~H.} \bibnamefont{Dederichs}},
  \bibinfo{author}{\bibfnamefont{V.}~\bibnamefont{Drchal}}, \bibnamefont{and}
  \bibinfo{author}{\bibfnamefont{J.}~\bibnamefont{Kudrnovsk\'{y}}},
  \bibinfo{journal}{Phys. Rev. B} \textbf{\bibinfo{volume}{65}},
  \bibinfo{pages}{064425} (\bibinfo{year}{2002}).

\bibitem[{\citenamefont{Butler et~al.}(2001)\citenamefont{Butler, Zhang,
  Schulthess, and MacLaren}}]{Butler:prb01}
\bibinfo{author}{\bibfnamefont{W.~H.} \bibnamefont{Butler}},
  \bibinfo{author}{\bibfnamefont{X.-G.} \bibnamefont{Zhang}},
  \bibinfo{author}{\bibfnamefont{T.~C.} \bibnamefont{Schulthess}},
  \bibnamefont{and} \bibinfo{author}{\bibfnamefont{J.~M.}
  \bibnamefont{MacLaren}}, \bibinfo{journal}{Phys. Rev. B}
  \textbf{\bibinfo{volume}{63}}, \bibinfo{pages}{054416}
  (\bibinfo{year}{2001}).

\bibitem[{\citenamefont{Mathon and Umerski}(2001)}]{Mathon:prb01}
\bibinfo{author}{\bibfnamefont{J.}~\bibnamefont{Mathon}} \bibnamefont{and}
  \bibinfo{author}{\bibfnamefont{A.}~\bibnamefont{Umerski}},
  \bibinfo{journal}{Phys. Rev. B} \textbf{\bibinfo{volume}{63}},
  \bibinfo{pages}{220403(R)} (\bibinfo{year}{2001}).

\end{thebibliography}

\end{document}